\documentclass[11pt]{article}
\usepackage{my_preamble}
\hypersetup{pageanchor=false}
\title{Scale Economies and Aggregate Productivity
}

\author{
Joel Kariel\thanks{Competition and Markets Authority, joel.kariel@cma.gov.uk} \and Anthony Savagar\thanks{University of Kent, a.savagar@kent.ac.uk. 
\newline
This research is funded under ESRC project reference ES/V003364/1.
\newline
\textbf{Disclaimer:} \textit{This work was produced using statistical data from ONS. The use of the ONS statistical data in
this work does not imply the endorsement of the ONS in relation to the interpretation or analysis of
the statistical data. This work uses research datasets which may not exactly reproduce National
Statistics aggregates.}
\newline
We thank seminar participants at UAB-IAE, Danmarks Nationalbank, Bank of Lithuania, GEA Tbilisi, Ghent Macro, CAED Penn State, IIOC Boston, Cardiff, CompNet, Nottingham, York, Durham Macro, EUI, Bank of Italy, MWM Clemson, University of Washington, RES 2023, EMF Bern 2023, Birmingham, St Louis Fed, Lancaster, King's, Bath, IFN Stockholm, EEA-ESEM 2022, IAAE 2022, CEF 2022, RES 2022, SNDE 2022, AMEF 2022, SES 2022, MMF 2022, Kent Firm Dynamics Workshop 2021, Exeter Macro Workshop 2022 and Bristol for their helpful comments. We thank the following people for feedback: Matthias Kehrig, Jan de Loecker, Jan Eeckhout, Mark Bils, Alex Monge, Max Gillman, Mark Wright, Omar Licandro, Julian Neira, Tom Schmitz, Petr Sedl\'{a}\v{c}ek, Danial Lashkari, John Morrow, Anthony Priolo, Riccardo Silvestrini and Kunal Sangani.
}
}
        
\date{\today}	
	
\begin{document}

\begin{titlepage}
\clearpage
\maketitle
\thispagestyle{empty}
\normalsize

\begin{abstract}
\onehalfspacing
We develop a theoretical framework to investigate the link between rising scale economies and stagnating productivity. Our model features heterogeneous firms, imperfect competition, and firm selection.  We demonstrate that scale economies generated by fixed costs have distinct impacts on aggregate productivity compared to those driven by returns to scale (slope of marginal cost). 
Using UK data, we estimate long-run increases in both fixed costs and returns to scale.
Our model implies that this should increase aggregate productivity through improved firm selection and resource allocation. However, increasing markups can offset the productivity gain. Higher markups cushion low-productivity firms' revenues, allowing them to survive, and constrain firm output, which limits exploitation of scale economies.
\\ [1em]
\\ [1em]
\end{abstract}

\end{titlepage}

\doublespacing
\hypersetup{pageanchor=true}

\section{Introduction}

Recent technological advances, such as cloud computing, digital platforms, and enterprise software, have increased firms' ability to scale production at lower cost. Yet over the last two decades, aggregate productivity growth in advanced economies such as the UK and US has stagnated. This coexistence of rising scale economies and weak productivity growth presents a puzzle. If firms can produce more efficiently at scale, why has aggregate productivity failed to accelerate?

This paper develops a theory linking firm-level scale economies to aggregate productivity. We show that rising scale economies -- whether generated by higher fixed costs or higher returns to scale in variable inputs -- should, ceteris paribus, raise aggregate productivity through stronger firm selection and more efficient division of resources. However, this productivity-enhancing channel can be substantially weakened, or offset, by rising markups. Higher markups relax firms’ zero-profit constraints, allowing low-productivity firms to survive.

We make three contributions. First, using UK firm-level data, we document a sustained increase in scale economies since the late 1990s, via both higher returns to scale and higher fixed costs. Second, we develop a tractable heterogeneous-firm model with monopolistic competition, endogenous entry, and scale economies that isolates how these forces affect aggregate productivity. Third, we show quantitatively that rising markups can reconcile increasing scale economies with stagnant aggregate productivity by weakening firm selection and distorting the allocation of labour across firms.

Underlying these contributions is a central insight that fixed costs and returns to scale, while observationally similar at the firm level, operate through distinct aggregate channels. Both reduce profits and lower the mass of active firms, strengthening selection and raising the average productivity of producers. However, the productivity gains from returns to scale depend on firms’ ability to expand output. When markups are high, demand constraints limit firm size, dampening the selection and allocation effects of increasing returns. In contrast, the impact of fixed costs on productivity is independent of markups under monopolistic competition, but extending to endogenous markups introduces a countervailing effect. With endogenous markups, low-productivity firms are cushioned by a rise in their markup when fixed costs increase, which dampens gains to aggregate productivity.

We decompose aggregate productivity into two components. The first is an extensive margin, which reflects the number of active firms and therefore the split of aggregate labour between production and overhead activities. The second is average productivity, which reflects the mean productivity of operating firms and is determined by selection at the entry margin. Rising scale economies strengthen selection and raise average productivity, but higher markups weaken this channel by allowing less productive firms to cover fixed costs. At the same time, higher markups increase the number of active firms, raising overhead labour and reducing the share of labour allocated to production.

We apply the model to the UK economy using estimated time series for returns to scale, fixed costs, and markups. We show that rising returns to scale, taken alone, would have generated substantial productivity gains. Absent the increase in markups, aggregate productivity would have been roughly 20 percent higher over the period we study. Once rising markups are taken into account, however, the productivity gains from scale economies are largely offset, generating dynamics consistent with the observed stagnation in UK productivity. 

Our paper abstracts from specific technologies, characterising them instead by their effects on fixed or marginal costs. Recent applied evidence supports the link between new technologies and scale economies. \textcite{Ganapati2025_AEJMicro} shows that information technology reduced marginal costs and increased markups in wholesale. \textcite{BloomGaricanoSadunVanReenen2014_ManagementScience} study new enterprise software that expands managerial span of control. \textcite{Syverson2019_JEP} highlights the rise of low-marginal-cost products, while \textcite{LashkariBauerBoussard2024_AER} link IT price declines to rising scale economies. 

Our results highlight the importance of market structure for the aggregate consequences of technological change. Technologies that raise scale economies need not translate into higher aggregate productivity if rising markups limit firm expansion and weaken competitive selection. More broadly, the paper shows that understanding productivity dynamics requires jointly accounting for scale economies, market power, and the extensive margin of firm activity.

\subsubsection*{Related Literature}

Our paper connects theory on the aggregate impacts of microeconomic production primitives, with the measurement of these features at the firm level. Recent work by \textcite{BilbiieMelitz2020_NBER, EdmondMidriganXu2023_JPE, BaqaeeFarhiSangani2023_Restud} demonstrates the importance of returns to scale for aggregate welfare. This work focuses primarily on external returns to scale (love of variety) that arise from aggregation. However, \textcite{BaqaeeFarhiSangani2023_Restud, BaqaeeFarhi2020_NBER} also note that returns to scale at the firm level magnify aggregate returns to scale. Similarly to our analysis, the effects of scale economies are smaller in efficient (low markup) economies. We combine theory with measurement to show that firm-level scale economies are quantitatively-relevant to replicate UK productivity dynamics.

In order to understand the consequences of rising market power, \textcite{DeLoeckerEeckhoutMongey2021_NBER} present a quantitative model with oligopolistic competition and fixed costs. This allows them to compare the role of technology on the supply-side versus competitive factors on the demand-side. 
We differ by focusing on analytical results to understand the supply-side mechanisms through which different technologies affect scale economies, and in turn aggregate productivity. We provide a partial equilibrium extension to endogenous markups which shows how the rise in markups from fewer firms can counteract gains in productivity.
Collectively, our papers advance the idea that to reconcile changing technologies on the supply side with stagnant aggregate performance, market power must increase on the demand side. 

Recent research in endogenous growth theory shows that changing technologies affect firm cost structures, which in turn explains stagnating growth.  \textcite{deRidder2024_AER} models intangible inputs as reducing marginal costs and raising fixed costs. Unlike us, the focus is the level of constant marginal costs, not the slope of marginal costs. \textcite{AghionBergeaudBoppartKlenowLi2023_ReStud} model a fixed cost that increases with the number of product lines, but as technology improves, the fixed cost becomes less sensitive to the number of products. Our paper differs from this research, which focuses on quantitative endogenous growth models with an important role for R\&D, and a main aim of replicating US stagnation facts. We present a parsimonious and tractable analysis based on firm entry to directly link the firm-level determinants of scale economies to aggregate productivity. Conceptually, this body of work, including our paper, contributes to the hypothesis that recent changes in technology have affected firm cost structures, and lead to important aggregate effects. To our knowledge, our work is the first to directly compare the effect of returns to scale and fixed costs, and formalise these effects of new technologies through the economies of scale channel. Informally, it is understood that these are two sources of scalable technologies and are hallmarks of intangible capital \parencite{HaskelWestlake2017_book}. We formalise that they both affect scale economies in the same way, but can lead to distinct aggregate productivity effects. 

Our model is a neoclassical growth model with heterogeneous firms based on \textcite{HopenhaynRogerson1993_JPE, RestucciaRogerson2008_RED, BarseghyanDiCecio2016_EER}. The model is similar to two-factor closed-economy versions of \textcite{Melitz2003_ecta, GhironiMelitz2005_QJE}.
We include firm production with fixed costs and returns to scale similar to the models of \textcite{Kim2004_JEDC, AtkesonKehoe2005_JPE, BartelsmanHaltiwangerScarpetta2013_AER, Kim2021_JIE}. \textcite{KehrigGao2021_wp} present a partial equilibrium industry model under perfect competition and focus on cross-industry variation in returns to scale and productivity dispersion. Similarly to our theory, they show a positive relationship between productivity and returns to scale across industries, whereby a rise in returns to scale leads to selection of more-productive firms. 

Several recent articles provide estimates of returns to scale in the US economy. \textcite{KehrigGao2021_wp} estimate slightly decreasing returns to scale in US manufacturing firms. Using similar US data, \textcite{RuzicHo2021_ReStat} find a decline in returns to scale from 1982 to 2007. Using Compustat data, \textcite{Chiavari2022_wp} documents rising returns to scale through production function estimation, and \textcite[][Figure 7]{deLoeckerEeckhoutUnger2020_QJE} documents increasing overhead cost shares as evidence of rising scale economies. \textcite{BaqaeeFarhiSangani2023_Restud} also document economies of scale in US firms.
\textcite{LashkariBauerBoussard2024_AER} find cost elasticity below one for French corporations, which implies economies of scale. 
For the UK economy, \textcite{Oulton1996_JIE,HarrisLau1998_OEP,GirmaGorg2002_CEPR} document constant or slightly decreasing returns to scale for manufacturing firms. Recent evidence for EU countries suggests returns to scale increasing over time \parencite{McAdamMeinenPapageorgiouSchulte2024_WP}, consistent with our UK findings.

The remainder of the paper is organised as follows. Section 2 documents trends in returns to scale, fixed costs, and aggregate productivity using UK data. Section 3 presents the model and equilibrium conditions. Section 4 discusses equilibrium properties, analyses comparative statics and introduces the aggregate productivity decomposition. Section 5 considers endogenous markups. Section 6 presents the quantitative analysis. Section 7 concludes. 

\section{Empirical Motivation}\label{sec:empirical_motivation}
We are motivated by the presence of rising scale economies at the firm level, while aggregate measures of productivity are stagnating. 
Although we use the UK as a motivating case, the mechanism we propose is not country-specific. The interaction between micro-level scale economies and aggregate productivity is likely to operate in other advanced economies experiencing similar slowdowns alongside the diffusion of scalable technologies such as IT/internet, cloud computing, and AI.\footnote{The supplementary appendix provides background on scale economies, defining them as the inverse cost elasticity and showing how they depend on returns to scale and fixed costs.}

\subsection{Productivity}
Figure~\ref{fig:TFP_growth} shows UK aggregate TFP growth over time. Aggregate productivity growth increases until 2007 but then declines and stagnates. This holds with the smoother aggregate TFP time series or with the more volatile weighted TFP estimates from firm-level data. Both data capture the UK `productivity puzzle', which has been widely studied in academic literature  \parencite{BarnettBattenChiuFranklineSebastia-Barriel2014_BOEQB,GoodridgeHaskelWallis2016_economica}, policy debate and popular discussion .
\begin{figure}[H]
    \centering
    \singlespacing
    \caption{UK TFP Growth, 1997 - 2019}
    \label{fig:TFP_growth}
    \begin{tikzpicture}
\begin{axis}[
            width=0.95\linewidth,
            height=7cm,
            x tick label style={
		    /pgf/number format/1000 sep=},
		    axis lines = left,
	        ylabel={Index (1997=1)},
	        xlabel=Year,
            xticklabel style={rotate=90},
        	enlarge y limits, 
	        enlarge x limits,
            table/col sep=comma,
            ymajorgrids=true,
            grid style={draw=gray!25},
            legend columns=1,
            legend style={at={(0.5,0.02)}, anchor=south west},
		    ]
\addplot[smooth, thick, blue!60, solid] table [x=year, y=aggTFP, col sep=comma] {./tikz_pgf/lnTFP_index_firmTFP_aggTFP.csv};
\addlegendentry{TFP (PWT)}

\addplot[smooth, thick, red!60, dashed] table [x=year, y=firmTFP, col sep=comma] {./tikz_pgf/lnTFP_index_firmTFP_aggTFP.csv};
\addlegendentry{TFP (Firm-level avg.)}

\end{axis}

\end{tikzpicture}
    \caption*{\footnotesize 
    TFP (PWT) is from the Penn World Table 10.01 \parencite{FeenstraInklaarTimmer2015_AER}, accessed from FRED: \href{https://fred.stlouisfed.org/series/RTFPNAGBA632NRUG}{Total Factor Productivity at Constant National Prices for United Kingdom (RTFPNAGBA632NRUG)}. TFP (Firm-level avg.) is the sales-weighted average of firm-level productivity residuals estimated following our production function estimation methodology \parencite{GandhiNavarroRivers2020_JPE}.}
\end{figure}

\subsection{Returns to Scale in Variable Inputs}

To measure returns to scale, we estimate firm-level production functions on UK data from the Annual Respondents Database (ARDx). The data covers 1997 to 2021, contains approximately 50,000 firms each year, 11 million workers, and two-thirds of gross value added. Firms report a range of production data, including gross output, value added, labour, materials, and investment.\footnote{In the appendix, we provide details about the data, data cleaning, deflation, capital construction, SIC code matching, and summary statistics.} We assume that we observe variable inputs, net of fixed costs.

We assume that each firm $\jmath$ has the following Cobb-Douglas production function
\[y_{\jmath t} = A_{\jmath t} k_{\jmath t}^{\beta_k} \ell_{\jmath t}^{\beta_\ell} m_{\jmath t}^{\beta_m} \]
where $y_t,  k_{\jmath t}, \ell_{\jmath t}, m_{\jmath t}$ are firm gross output and inputs of capital, labour and intermediate inputs. $A_{\jmath t}$ is a measure of firm-level Hicks neutral technology which we do not observe. Our aim is to estimate the $\beta_k, \beta_\ell$ and $\beta_m$ parameters which represent output elasticities. The sum of these output elasticities is returns to scale in variable inputs.

Production function estimation suffers from omitted variable bias. The bias occurs because the input variables are correlated with the unobserved firm-level efficiency term. There are various methods to address this problem \parencite{OlleyPakes1996_ecta, LevinsohnPetrin2003_ReStud, AckerbergCavesFrazer2015_ecta, GandhiNavarroRivers2020_JPE}. We follow \textcite{GandhiNavarroRivers2020_JPE} which non-parametrically identifies elasticities with a gross output production function, even without exogenous sources of variation (e.g. prices). 

Since we estimate Cobb-Douglas production functions, we obtain a single, time-invariant, coefficient for each input in the production function. This is computed at different levels of sectoral disaggregation, yielding sector-specific estimates of returns to scale. These sectoral estimates are combined with firm-level data, allowing us to compute trends over time and analyse returns to scale across the firm distribution. Finally, we estimate production functions over different time periods to permit more flexibility in the underlying production technology over time.

Figure \ref{fig:RTS_GNR} shows estimated returns to scale across firms in the UK using the estimation methodology of \textcite{GandhiNavarroRivers2020_JPE}, allowing production functions to vary across sectors. There is clear rising returns to scale from 1997 to 2021. 
\begin{figure}[H]
    \centering
    \singlespacing
    \caption{UK RTS, 1997 - 2021}
    \label{fig:RTS_GNR}
    \begin{tikzpicture}

\begin{axis}[
            width=0.95\linewidth,
            height=7cm,
            x tick label style={
		    /pgf/number format/1000 sep=},
		    axis lines = left,
	        ylabel={Returns to Scale $(\nu)$, Levels},
	        xlabel=Year,
            xticklabel style={rotate=90},
        	enlarge y limits, 
	        enlarge x limits,
            table/col sep=comma,
            ymajorgrids=true,
            grid style={draw=gray!25},
            legend columns=1,
            legend style={at={(0.5,0.02)}, anchor=south west},
		    ]

\addplot[smooth, very thick] table [x=year, y=GNR_RTS, col sep=comma] {./tikz_pgf/rts_gnr_1997_2021.csv};

\end{axis}

\end{tikzpicture}
    \caption*{\footnotesize Returns to scale (RTS) are measured as the sum of coefficients from a Cobb–Douglas gross-output production function estimated using the methodology of \textcite{GandhiNavarroRivers2020_JPE}. We obtain time-varying RTS by estimating these coefficients at the 2-digit industry level, merging them to firm-level data, and computing sales-weighted annual averages.}
\end{figure}
However, Figure \ref{fig:RTS_GNR} only allows the production technology to differ \textit{across} sectors, not \textit{within} sectors over time. As an alternative, we do both: estimating sector-specific returns to scale on 5-year windows. The distribution of returns to scale across 2-digit SICs is shown in Figure \ref{fig:rts2d_periods_kdensity}, showing a rightwards shift from the late 1990s to the most recent data.
\begin{figure}[H]
    \centering
    \caption{Distribution of returns to scale across 2-digit sectors over time}
    \label{fig:rts2d_periods_kdensity}
    \includegraphics[width=.9\textwidth]{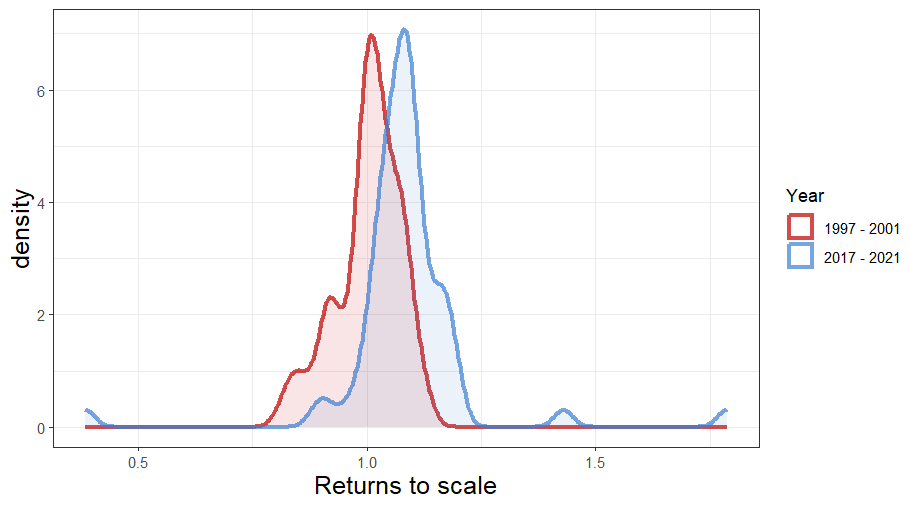}
\caption*{\footnotesize Returns to scale (RTS) are estimated as the sum of output elasticities from a Cobb--Douglas gross-output production function following \textcite{GandhiNavarroRivers2020_JPE}. Estimates are obtained at the 2-digit SIC level using rolling 5-year windows and are shown as kernel density distributions across sectors. The rightward shift of the distribution over time indicates a broad-based increase in returns to scale across UK industries.}
\end{figure}

In the appendix, we provide the underlying output elasticities and returns to scale estimates at the industry level (2-digit and 3-digit) on average and across time periods.
\subsection{Fixed Cost Share}

An alternative contributor to firm-level scale economies is the fixed cost share. In Figure \ref{figure:adshare_desc_stat}, we use two complementary proxies to capture the evolution of fixed costs.

The first proxy is the share of administration expenses in firm revenue, following \textcite{deLoeckerEeckhoutUnger2020_QJE}. The second is the share of employment in \emph{Administrative and Support Service Activities} (SIC code N).\footnote{\href{https://www.ons.gov.uk/employmentandlabourmarket/peopleinwork/earningsandworkinghours/datasets/averageweeklyearningsbyindustryearn03}{ONS EARN03 employment weights}.} The latter maps more closely to our model, in which fixed costs take the form of labour overheads that are required for operation but do not scale proportionally with output.

Both measures display a sustained upward trend over time, indicating rising overhead costs at the firm level. Administration expenses in UK company accounts capture costs that are not directly related to production, manufacturing, or sales activities. While this measure is an imperfect proxy for fixed costs, its secular rise is consistent with technological changes that increase non-production inputs such as management and compliance. Further details on data construction and definitions are provided in the Appendix.

Taken together, the accounts-based and labour-based measures offer complementary perspectives: one captures the cost share of overheads, while the other captures the allocation of labour away from production and toward non-production activities. Both point to an increase in fixed costs, reinforcing the interpretation of rising firm-level scale economies.

\begin{figure}[H]
    \centering
    \singlespacing
    \caption{Fixed Cost Growth from 2004--2023}
    \label{figure:adshare_desc_stat}
    \begin{tikzpicture}
\begin{axis}[
            width=0.95\linewidth,
            height=7cm,
            x tick label style={
		    /pgf/number format/1000 sep=},
		    axis lines = left,
	        ylabel= {Index (2004=1)},
	        xlabel={Year},
            xticklabel style={rotate=90},
        	enlarge y limits, 
	        enlarge x limits,
            table/col sep=comma,
            xtick={2004,2006,2008,2010,2012,2014,2016,2018,2020,2022,2024},
            ymajorgrids=true,
            grid style={draw=gray!25},
            legend columns=1,
            legend style={at={(0.5,0.02)}, anchor=south west},
		    ]

\addplot[smooth, thick, blue!60, dashed] table [x=year, y=fixed_cost_idx, col sep=comma] {./tikz_pgf/adshare_desc_stat.csv};
\addlegendentry{Admin. Expenses Share}

\addplot[smooth, thick, purple!60] table [x=year, y=admin_support_idx, col sep=comma] {./tikz_pgf/adshare_desc_stat.csv};
\addlegendentry{Admin. Labour Share}

\end{axis}

\end{tikzpicture}
    \caption*{\footnotesize The figure shows the growth in (i) the median share of `Administration Expenses' in `Turnover' for UK firms from BvD FAME, and (ii) the share of employment in `Administrative and Support Service Activities' from the ONS, EARN03 employment weights.}
\end{figure}

Combined with the evidence of rising returns to scale in variable inputs and stagnant aggregate TFP, these patterns highlight a tension: technologies appear to increase firms’ scalability, yet aggregate productivity fails to improve. Resolving this tension requires a framework in which firm size, entry, fixed costs, and markups jointly determine selection, resource allocation, and aggregate outcomes. Section~\ref{sec:model} develops such a model.
\section{Model \& Equilibrium}\label{sec:model}
The household side of the model follows a neoclassical growth setup. The production side of the economy has firm entry and exit, monopolistic competition, and production functions that have fixed costs and returns to scale. 

\subsection{Households}
A representative household maximizes lifetime utility subject to a budget constraint
\begin{align} \label{eq:hh_problem}
\max_{\{C_t, K_{t+1}\}^\infty_{t=0}} \quad & \sum_{t=0}^\infty \beta^t \frac{C_t^{1-\sigma} - 1}{1 - \sigma}, \quad \beta \in (0,1), 
\nonumber
\\
\textrm{s.t.} \quad & C_t + I_t = r_t K_t + w_t L^{\mathrm{s}} + \Pi_t + T_t
\\
& I_t = K_{t+1} - (1-\delta) K_t.
\end{align}
Households own all firms in the economy and receive profits $\Pi_t$. $T_t$ is a lump sum transfer from the government that will be equal to the entry fees paid by the firms. Households supply a fixed amount of labour that is not time-varying, we normalize this to one:
\begin{equation}
    L^{\mathrm{s}}=1.
\end{equation}
Households own the capital stock and rent it to firms at a rental rate $r_t$, hence the capital investment decision is part of the household problem. 
The household optimization problem satisfies the following condition
\begin{equation}
    \left( \frac{C_{t+1}}{C_t} \right)^\sigma = \beta (r_{t+1} + (1 - \delta)).
\end{equation}
plus a transversality condition and the resource constraint.

\subsection{Firms}
The key modelling feature is that scale economies arise jointly from fixed costs and returns to scale in variable inputs, while market power limits the extent to which these scale economies translate into firm size and aggregate productivity.

\subsubsection{Final goods producer}
The final goods aggregator is a generalised CES:
    \begin{align}\label{eq:CES_aggregator}
    Y_t = N_t \left[ \frac{1}{N_t} \int_0^{N_t} y_t(\imath)^{\frac{1}{\mu}} d\imath \right]^\mu.
    \end{align}
There are $N_t$ intermediate producers on the interval $\imath \in (0,N_t)$. The parameter $\mu \geq 1$ captures product substitutability.\footnote{Perfectly substitutable products ($\mu=1$) are admissible when intermediate producers face fixed costs and decreasing returns to scale ($\phi>0$ and $\nu\in(0,1)$). This corresponds to perfect competition, in which firms take prices as given and operate at the minimum of their average cost curves under perfectly elastic demand.} The aggregator has constant returns to scale.\footnote{Under a CES aggregator, the pre-multiplier is $N_t^{\mu}$ and offsets the $1/N_t$ term inside the brackets. This introduces additional increasing returns in aggregation, which we abstract from.}
Under monopolistic competition, $\mu$ also governs the equilibrium markup, with a higher value corresponding to greater market power and a higher price–cost margin.

The maximization problem of the final goods producer is
\begin{align*}
\Pi^F_t = \max_{y_t(\imath)} \quad &  Y_t - \int_0^{N_t} p_t(\imath) y_t(\imath) d\imath
\end{align*}
subject to \eqref{eq:CES_aggregator}.
The firm is infinitesimal so firm level output does not affect $Y_t$. The first-order condition with respect to $y_t(\imath)$ gives the inverse-demand for a firm
\begin{align} \label{eq:inverse_demand_function}
     p_t(\imath) = \left( \frac{N_t y_t(\imath)}{Y_t}\right)^{\frac{1-\mu}{\mu}}.
\end{align}

\subsubsection{Intermediate goods producer}

An entrant pays the entry cost $\kappa$ and draws productivity $\jmath\sim U(0,1)$, which maps into $A(\jmath)$. The firm then decides whether to produce, incurring a fixed overhead cost; otherwise it remains inactive. All firms exit after one period. A firm with productivity $\jmath$ has the following production function
\begin{equation} \label{eq:firm_output}
y_t(\jmath)
=
A(\jmath)\bigl[k_t(\jmath)^{\alpha}\ell_t(\jmath)^{1-\alpha}\bigr]^{\nu},
\end{equation}
where $0<\alpha<1$ is the capital share in variable costs and $\nu>0$ governs returns to scale in variable inputs and the slope of marginal cost. Returns are decreasing, constant, or increasing for $\nu\lessgtr 1$. Production labour satisfies
\begin{equation} \label{eq:production_labour}
\ell_t(\jmath)=\ell_t^{\mathrm{tot}}(\jmath)-\phi,
\end{equation}
with $\phi$ a fixed overhead cost. Both $\nu$ and $\phi$ determine scale economies, measured by the inverse cost elasticity.

The firm solves
\begin{equation}
\max_{k_t(\jmath),\ell_t(\jmath)}
\; p_t(\jmath)y_t(\jmath)
- r_t k_t(\jmath)
- w_t\bigl(\ell_t(\jmath)+\phi\bigr),
\end{equation}
subject to \eqref{eq:firm_output} and inverse demand \eqref{eq:inverse_demand_function}. Optimality implies constant factor shares in revenue,
\begin{align}\label{eq:firm_profit_problem_focs}
\frac{r_t k_t(\jmath)}{p_t(\jmath)y_t(\jmath)} &= \frac{\nu}{\mu}\alpha,
&
\frac{w_t \ell_t(\jmath)}{p_t(\jmath)y_t(\jmath)} &= \frac{\nu}{\mu}(1-\alpha),
\end{align}
so the variable–cost share in revenue is $\nu/\mu$, with capital and labour shares in \textit{variable costs} $\alpha$ and $1-\alpha$, respectively. The residual revenue share $1-\nu/\mu$ corresponds to profits plus fixed costs.

A necessary condition for profit maximization is $\nu<\mu$, ensuring an interior $MR=MC$ solution. We additionally impose $\alpha\nu<1$ to guarantee concavity of aggregate capital demand. Together,
\begin{equation*}
\nu < \min\!\left\{\frac{1}{\alpha},\,\mu\right\}.
\end{equation*}
Hence, higher markups and lower capital shares in variable costs permit greater returns to scale in variable inputs.

\subsubsection{Relative firm size}

Firm output, revenue, and input demands are proportional to productivity, transformed by the difference between markups and returns to scale. In particular,
\[
y(\jmath)^{1/\mu},\; p(\jmath)y(\jmath),\; k(\jmath),\; \ell(\jmath)
\;\propto\;
A(\jmath)^{\frac{1}{\mu-\nu}}.
\]
Therefore,  $A(\jmath)^{\frac{1}{\mu-\nu}}$ is the effective productivity variable which determines heterogeneity in economic variables, and which disciplines the firm size distribution. Higher markups or lower returns to scale flatten the mapping from productivity to firm size, compressing the firm size distribution. 
Intuitively, higher markups cushion low-productivity firms’ revenues while constraining the expansion of high-productivity firms, weakening reallocation toward the upper tail.
\subsubsection{Zero-profit firm}
We assume there is a threshold productivity draw $J_t \in (0,1)$ characterised by zero profits, which yields threshold technology $\underbar A_t$. If a firm receives a productivity draw below the threshold productivity level they would make negative profits from production. Consequently, they prefer to produce zero and make zero profits. Therefore we define profits and characterise the threshold productivity as follows:
\begin{align}
     \pi_t(\jmath) &= p_t(\jmath)y_t(\jmath) - r_t k_t(\jmath) - w_t(\ell_t(\jmath)+\phi)
     \\
    \pi_t(J_t) &= 0.
\end{align}
 
Combining the profit condition with factor prices yields a helpful reduced-form relationship: 
\begin{equation}\label{eq:rf_profit}
    \pi_t(\jmath) = \left(1 - \frac{\nu}{\mu}\right)p_t(\jmath)y_t(\jmath) - w_t\phi.
\end{equation}
Furthermore, with the zero-profit condition and with the ratio of revenues to scaled productivity:
\begin{align}\label{eq:profit_relative_As}
    \pi_t(\jmath) = \phi w_t \left[\left(\frac{A(\jmath)}{\underbar A_t} \right)^{\frac{1}{\mu-\nu}} -1 \right].
\end{align}

\subsubsection{Free Entry}
All firms die after one period. A firm produces if it makes positive profits, hence firm value is given by
\begin{equation}\label{eq:v}
    v_t(\jmath) = \max \{\pi_t(\jmath),0\}.
\end{equation}
We assume a free entry condition which implies that the unconditional expected value from entering equals to the entry cost $\kappa$:
\begin{equation}
 \mathbb{E} [v_t(\jmath)] = \kappa. \label{eq:free_entry_1}
\end{equation}
The cost of entry $\kappa$ is denominated in consumption units and is rebated to households in a lump-sum.
Combining \eqref{eq:v} and \eqref{eq:free_entry_1} with our reduced-form profit expression \eqref{eq:profit_relative_As} yields:
\begin{equation} \label{eq:free_entry_2}
      \phi w_t (1-J_t) \left[ \left(\frac{\hat{A}_t}{\underbar{A}_t}\right)^{\frac{1}{\mu - \nu}} - 1\right] = \kappa.
\end{equation}
This shows that (unconditional) expected profits from being active multiplied by the probability of being active $1-J_t$ equals the entry cost.
We have defined the power mean of technology, conditional on being active, as
\begin{equation}\label{eq:technology_power_mean}
    \hat{A}(J_t) \equiv \mathbb{E} \left[ \left. A(\jmath)^{\frac{1}{\mu-\nu}} \right| \jmath > J_t \right]^{\mu - \nu} = \left[ \frac{1}{1-J_t} \int_{J_t}^1 A(\jmath)^{\frac{1}{\mu-\nu}} \; d \jmath \right]^{\mu-\nu}.
\end{equation}
The power mean is a weighted average of firm-level productivity.\footnote{The term $\hat{A}(J_t)$ generalizes \textcite[eq. 7][p. 1700]{Melitz2003_ecta} and \textcite[eq. 31][p. 10]{ColciagoSilvestrini2022_EER}. This term is equivalent to these papers if $\nu=1$ and the markup is expressed in terms of elasticities of substitution between goods. 
} 
Throughout, we work in quantile $\jmath$-space, which makes selection transparent.

\subsection{Aggregation}

A mass $E_t$ of firms enters at date $t$. Entrants are indexed by $\jmath\in(0,1)$ and draw productivity upon entry. Firms with $\jmath>J_t$ choose to produce, implying a mass of active firms
\begin{equation}
N_t=E_t(1-J_t).
\end{equation}
Throughout, integrals over active firms indexed by $\imath\in(0,N_t)$ are equivalent to integrals over entrants indexed by $\jmath\in(J_t,1)$ and scaled by $E_t$. For example, $
N_t
\;=\;
\int_0^{N_t} d\imath
\;=\;
E_t\int_{J_t}^1 d\jmath
\;=\;
E_t(1-J_t).$

Aggregate capital and labour are
\begin{align}
K_t
&=
\int_0^{N_t} k_t(\imath)\, d\imath,
\label{eq:agg_K}
\\
L_t
&=
\int_0^{N_t}\!\big[\ell_t(\imath)+\phi\big]\, d\imath.
\label{eq:agg_L}
\end{align}
Define the share of labour allocated to production as
\begin{equation}
u_t
\;\equiv\;
\frac{\int_0^{N_t} \ell_t(\imath)\, d\imath}{L_t},
\label{eq:u}
\end{equation}
so that the share of labour devoted to overhead activities is
\begin{equation}
1-u_t
=
\frac{N_t\phi}{L_t}.
\label{eq:1-u}
\end{equation}

We can express aggregate output as:
\begin{equation}
Y_t
=
N_t^{\,1-\nu}\,
\hat A_t
\Big[
K_t^{\alpha}(u_t L_t)^{1-\alpha}
\Big]^{\nu}.
\label{eq:Y}
\end{equation}
Taking $N_t^{-\nu}$ inside the brackets clarifies that aggregate output is the sum across $N_t$ symmetric firms, each with average technology $\hat{A}_t$. This expression is homogeneous of degree one in capital, production labour, and the number of active firms. Holding $N_t$ fixed, aggregate output is homogeneous of degree $\nu$ in capital and production labour.

The wage, rental rate on capital and zero-profit condition are
\begin{align}
    r_t &= \alpha \frac{\nu}{\mu} \frac{Y_t}{K_t} \label{eq:agg_r}
    \\
    w_t &= (1-\alpha) \frac{\nu}{\mu} \frac{Y_t}{u_t L_t} \label{eq:agg_w}
    \\
    \frac{w_t}{Y_t}N_t \phi &
    = \left(1-\frac{\nu}{\mu}\right)\left(\frac{\underbar{A}_t}{\hat{A}_t}\right)^{\frac{1}{\mu - \nu}}.  \label{eq:agg_zero_profit}
\end{align}

\subsection{Government Budget Constraint and Resource Constraints}
The resource constraint is
    \begin{equation}\label{eq:agg_resource_constraint}
        Y_t = C_t + I_t.
    \end{equation}
The government rebates entry fees to households. The government budget constraint equates taxes to government expenditure
    \begin{equation}
        T_t = E_t \kappa. \label{eq:gov_bc}
    \end{equation}
Profits and labour markets clear:
    \begin{align}
        \Pi_t &= \Pi^F_t \label{eq:profit_clearing}
        \\
        L_t &= L^s. \label{eq:L_market_clearing}
    \end{align}
Aggregate profits received by the household from owning firms equate to profits earned by the final goods producer. The profits are zero in equilibrium. Labour demanded by the firm equates to labour supplied by the household which is normalised to 1.
        
\subsection{Equilibrium Definition}

An equilibrium is a sequence of prices $\{r_t,w_t\}_{t=0}^\infty$; firm-level input choices $\{\ell_t(\jmath),k_t(\jmath)\}_{t=0}^\infty$; firms’ operating decisions and the associated measures of entry and active firms $\{E_t,N_t\}_{t=0}^\infty$; and aggregate allocations $\{C_t,K_{t+1}\}_{t=0}^\infty$, such that:
\begin{enumerate}
    \item households choose $\{C_t,K_{t+1}\}$ to solve the household problem \eqref{eq:hh_problem};
    \item firms optimally choose whether to produce and, if active, their factor demands, satisfying \eqref{eq:firm_profit_problem_focs};
    \item the free-entry condition holds \eqref{eq:free_entry_1};
    \item markets clear for labour \eqref{eq:agg_L}, capital \eqref{eq:agg_K}, goods \eqref{eq:agg_resource_constraint}, and profits \eqref{eq:profit_clearing};
    \item the government budget constraint is satisfied \eqref{eq:gov_bc}.
\end{enumerate}

\section{Equilibrium Properties and Comparative Statics} \label{sec:theoretical_analysis}

Section~\ref{sec:model} defined the equilibrium. This section derives its key properties and comparative statics, focusing on how fixed costs, returns to scale, and markups shape firm behaviour and aggregate outcomes.
We establish three results. First, without imposing a productivity distribution, aggregate factor shares are pinned down by technology and markups, while selection affects only the division of labour between production and overhead. Second, under Pareto productivity, returns to scale strengthen selection while markups weaken it, generating clean comparative statics for firm mass and labour shares. Third, in steady state, aggregate productivity responds to fixed costs and returns to scale through distinct extensive margin and average productivity (selection) channels. Markups have no effect on the fixed cost response, but dampen the returns to scale response.

\subsection{Aggregate Objects}\label{sec:agg_objects}
Combining the equilibrium conditions implies constant aggregate output shares for aggregate capital and production labour, and therefore a constant variable cost share. These shares are accounting identities and do not depend on firm selection or the productivity distribution. Since they are constant, we drop time subscripts.
\begin{equation}
s_{K}
=
\frac{r_t K_t}{Y_t}
=
\frac{\alpha\nu}{\mu},
\qquad
s^{\mathrm{prod}}_{L}
=
\frac{w_t u_t L_t}{Y_t}
=
\frac{(1-\alpha)\nu}{\mu}
, \qquad s_{VC}=
\frac{r_t K_t + w_t u_t L_t}{Y_t} = \frac{\nu}{\mu}
.
\end{equation}
Hence, the aggregate capital share in variable costs is given by $\alpha$ and the production labour share in variable costs is given by $1-\alpha$. Define the productivity gap between average firm productivity and the threshold productivity as
\begin{equation}
\Gamma_t(\underline{A}_t)
\;\equiv\;
\frac{\hat{A}_t}{\underline{A}_t}
\;\ge\; 1 .
\end{equation}
An increase in $\Gamma_t$ implies that the average productivity of active firms lies further above the cut-off, reflecting stronger selection. 

While factor shares are pinned down mechanically, selection affects aggregates through the allocation of labour between production and overhead activities. The aggregate equilibrium conditions \eqref{eq:agg_r}, \eqref{eq:agg_w}, and \eqref{eq:agg_zero_profit} imply that the share of labour allocated to production is:
\begin{equation}
u_t
=
\left[
1
+
\frac{1}{1-\alpha}
\left(
\frac{\mu}{\nu}-1
\right)
\Gamma_t^{-1/(\mu-\nu)}
\right]^{-1}.
\end{equation}
The expression for $u_t$ is a reduced-form identity. Its economic content lies in how selection ($\Gamma_t$) moves $u_t$ in equilibrium. Production-labour allocation is increasing in selection ($\Gamma_t$), and since the number of operating firms behaves inversely as $N_t = (1-u_t)/\phi$, it follows that $N_t$ is decreasing in $\Gamma_t$. 
\[\frac{\partial u_t}{\partial \Gamma_t} >0, \qquad \frac{\partial N_t}{\partial \Gamma_t} <0.\]
Other aggregate shares depend on selection ($\Gamma_t$) through production labour $u_t$:
\begin{align}
s_{\phi,t}
&=
\frac{w_t N_t \phi}{Y_t}
=
\frac{(1-\alpha)\nu}{\mu}\,
\frac{1-u_t}{u_t},
\\
s_{L,t}
&=
\frac{w_t L_t}{Y_t}
=
\frac{(1-\alpha)\nu}{\mu}\,
\frac{1}{u_t},
\\
s_{\pi,t}
&=
1 - s_{K} - s_{L,t}.
\end{align}
The profit share is the residual of the total cost share $s_{TC,t} = s_{K} + s_{L,t} = s_{VC} + s_{\phi,t}$.
Importantly, $\Gamma_t$ influences aggregate shares only through the labour-allocation margin. Increasing production-labour allocation, decreases the overhead-labour allocation, decreases the total labour share, decreases the total cost share, and increases its residual: the profit share. 

The results show that selection is a key determinant aggregate shares:
\begin{equation}
\frac{\partial s_{\phi,t}}{\partial \Gamma_t} < 0,
\qquad
\frac{\partial s_{L,t}}{\partial \Gamma_t} < 0,
\qquad
\frac{\partial s_{\pi,t}}{\partial \Gamma_t} > 0.
\end{equation}
As selection increases, more total labour goes to production ($u_t \to 1$) and the overhead cost component disappears ($1-u_t \to 0$). Consequently, the overhead share decreases to zero, the total labour share decreases towards the production-labour share and the profit share increases. Economically, greater selection implies a larger productivity gap reflecting a greater density of large, high-productivity firms relative to the cut-off.
Since all firms pay the same fixed cost, this decreases the aggregate fixed-cost share and, since production-labour and capital shares are fixed, the residual profit share increases.

Changes in structural parameters, such as $\{\nu, \mu, \phi\}$, impact aggregate shares directly and indirectly through $u_t$ which in turn depends on a direct effect and an indirect effect from $\Gamma_t$.\footnote{Formally, for any aggregate object $X_t \in \{s_{\phi,t}, s_{L,t}, s_{\pi,t}\}$, the response to a change in $x \in \{\nu,\mu,\phi\}$ is
\[
\frac{d X_t}{d x}
=
\frac{\partial X}{\partial x}
\;+\;
\frac{\partial X}{\partial u_t}
\;
\underbrace{
\Bigg[
\frac{\partial u_t}{\partial x}
\;+\;
\frac{\partial u_t}{\partial \Gamma_t}
\;
\overbrace{
\Big(
\frac{\partial \Gamma_t}{\partial x}
\;+\;
\frac{\partial \Gamma_t}{\partial \underline{A}_t}
\;
\frac{d \underline{A}_t}{d x}
\Big)
}^{\displaystyle d \Gamma_t/d x}
\Bigg]
}_{\displaystyle d u_t/d x} .
\]
This describes (distribution-free and out-of-steady-state) comparative statics of the static equilibrium conditions.
To pin down $d\Gamma_t/dx$, one must specify the productivity distribution. To pin down $d\underline{A}_t/dx$, one must impose steady-state. 
In a supplementary appendix, we derive these results. They are ambiguous without distributional assumptions. 
}
To get precise results, we derive these responses for the Pareto case below.

Lastly, we note that neither factor shares nor labour allocation depend on $\phi$. The fixed cost affects the equilibrium number of active firms but does not influence firm-level input demands. Conditional on being active, all firms make identical production decisions regardless of $\phi$. The fixed cost simply scales the mass of firms taking those decisions. 

\subsubsection{Aggregate Productivity}
Aggregate measured productivity in this model is shaped by two components. The first is average productivity, which reflects selection of active firms. The second is the extensive margin (mass of firms), which captures how aggregate labour is distributed across active firms. Because every firm duplicates the fixed overhead cost and potentially operates under non-constant returns, the number of firms determines both how much labour is absorbed by overhead and the returns to scale effect of dividing aggregate labour across firms.

We can write aggregate output in Cobb-Douglas form as:
\begin{align} \label{eq:Y_CobbDouglas}
    Y_t &= TFP_t K_t^{\alpha \nu}L_t^{1-\alpha \nu}
    \\
    \text{where,} \quad TFP_t &\equiv \left(\frac{N_t}{L_t}\right)^{1-\nu} \left( 1 - \frac{N_t \phi}{L_t} \right)^{(1-\alpha)\nu} \hat{A}_t 
    \\
    &
    = \left(\frac{1-u_t}{\phi}\right)^{1-\nu} u_t^{(1-\alpha)\nu} \hat{A}_t  \label{eq:TFP}
\end{align}
We define \textit{measured} aggregate TFP as aggregate output net of aggregate capital and labour inputs.
\footnote{It is the residual from the following measurement equation $\ln TFP_t = \ln Y_t - \alpha \nu \ln K_t - (1-\alpha) \nu \ln L_t$, using data on aggregate output, capital and labour. It differs from the conventional Solow Residual because the coefficients do not correspond to factor shares due to the markup.} 
This object reflects both selection and allocative forces.
We then decompose TFP into an extensive margin (number of firms) and average productivity component:
\begin{equation}
    TFP_t = \underbrace{\Omega_t}_{\text{extensive margin}} \times \underbrace{\hat{A}_t}_{\text{avg. prod.}}, \quad \text{where} \; \Omega_t \equiv  \underbrace{\left(\frac{N_t}{L_t}\right)^{1-\nu}}_{\text{Scale effect}} \times \underbrace{\left( 1 - \frac{N_t \phi}{L_t} \right)^{(1-\alpha)\nu}}_{\text{Resource duplication}}. 
\end{equation}
The extensive margin captures the negative effect of more firms duplicating fixed costs, and the scale effect of dividing aggregate labour across firms, which depends on returns to scale $\nu \gtreqless 1$.\footnote{With increasing returns ($\nu>1$), the extensive margin $\Omega_t$ falls in $N_t$ since concentration exploits scale and avoids fixed-cost duplication. With decreasing returns ($\nu<1$), $\Omega_t$ is concave and maximised at $N^\ast=(1/\phi),(1-\nu)/(1-\alpha\nu)$ (for $L_t=1$). In our Pareto case, equilibrium yields excess entry ($N>N^\ast$) due to fixed-cost duplication. $N$ is a constant equilibrium outcome, not a steady-state outcome.} 
Average productivity depends on the distribution of firm productivities conditional on entry. It is determined by selection.
This separation parallels the distinction emphasised by \textcite{Hopenhayn2014_ARE}, who shows that aggregate productivity depends jointly on the distribution of firm productivities and on the equilibrium number of firms. \textcite[Eq. 17]{JaimovichTerryVincent2023_NBER} provide a similar decomposition.

\subsection{Equilibrium with Pareto Distribution} \label{sec:Pareto_equilibrium}

To obtain closed-form expressions for the key aggregates, we assume that firm-level productivity is Pareto distributed. We relegate a full presentation of equilibrium conditions under Pareto to the appendix. 

Changes in structural parameters affect aggregate variables through a direct effect and through selection. For the Pareto distribution, we can isolate the effect on selection without needing characterise steady state because selection (the productivity gap) is constant: $\Gamma_t(\underbar{A}) = \Gamma$.

\begin{lemma}[Firm selection behaviour under Pareto]\label{lem:Gamma_derivatives} Given a Pareto distribution for productivity, selection is independent of fixed overhead costs, increasing in returns to scale and decreasing in the markup:
\begin{align*}
\frac{\partial \Gamma}{\partial \phi} = 0,\quad
\frac{\partial \Gamma}{\partial \nu} > 0,\quad
\frac{\partial \Gamma}{\partial \mu} < 0.
\end{align*}
\end{lemma}
\begin{proof}
    See Appendix~\ref{app:Pareto_CS_proofs}.
\end{proof}
The proof formalises this intuition. Higher returns to scale strengthen selection by disproportionately increasing the profitability of high-productivity firms, whereas higher markups weaken selection by cushioning low-productivity firms’ revenues and allowing them to survive.

\begin{proposition}[Comparative statics under Pareto]
\label{prop:Pareto_CS}
The comparative statics are as follows:
\begin{enumerate}[(i)]
    \item \emph{Production labour share.}
    \[
        \frac{d u}{d\phi} = 0,
        \qquad
        \frac{d u}{d\nu} > 0,
        \qquad
        \frac{d u}{d\mu} < 0.
    \]
    A higher fixed cost does not affect the fraction of labour in production; higher returns to scale increase it; a higher markup reduces it.

    \item \emph{Number of active firms.}
    \[
        \frac{dN}{d\phi} < 0,
        \qquad
        \frac{dN}{d\nu} < 0,
        \qquad
        \frac{dN}{d\mu} > 0.
    \]
    The mass of active firms falls with fixed costs and returns to scale and rises with the markup.

    \item \emph{Aggregate labour share.}
    \[
        \frac{d s_L}{d\phi} = 0,
        \qquad
        \frac{d s_L}{d\nu}  < 0,
        \qquad
        \frac{d s_L}{d\mu}  > 0.
    \]
    The labour share is invariant to $\phi$, decreasing in returns to scale $\nu$, and increasing in the markup $\mu$.
    \item \emph{Aggregate profit share.}
    \[
        \frac{d s_\pi}{d\phi} = 0,
        \qquad
        \frac{d s_\pi}{d\nu}  = 0,
        \qquad
        \frac{d s_\pi}{d\mu}  < 0.
    \]
    The profit share is invariant to $\phi$ and $\nu$, and decreasing in the markup $\mu$.
\end{enumerate}
\end{proposition}
\begin{proof}
See Appendix~\ref{app:Pareto_CS_proofs}.
\end{proof}

As explained in Section~\ref{sec:agg_objects}, the results relating to $\phi$ changes and invariance hold regardless of distributional assumptions or the imposition of steady-state. The Pareto distribution is widely used in firm dynamics models as a proxy of the firm size distribution. The scale invariance property yields precise comparative statics and a clear intuition for the mechanisms.  
Truncating a Pareto distribution preserves its shape:
average productivity among active firms is proportional to the cutoff,
$\hat A = \Gamma \underline A$, where $\Gamma = \left( \frac{\vartheta(\mu-\nu)}{\vartheta(\mu-\nu) - 1} \right)^{\mu-\nu}$ is a constant. The constant depends on the Pareto tail parameter  $\vartheta$ and, more consequentially, the effective tail parameter $\vartheta^\ast\equiv \vartheta(\mu-\nu)$ that determines the distribution of economic variables. As a result, changes in
structural parameters affect $\Gamma$ only through how the Pareto productivity distribution maps into economic variables, rather than through general-equilibrium feedbacks operating via the cutoff $\underline A$ itself.

Lemma~\ref{lem:Gamma_derivatives} shows that returns to scale and markups have
opposite effects on selection. Higher returns to scale amplify differences in
firm size and profitability across productivity draws, raising the average productivity relative to the cut-off (strong selection $\Gamma \uparrow$). Higher markups compress
the mapping from productivity to firm size and profits, raising the density of firms around the cut-off (weak selection $\Gamma \downarrow$). Markups increase revenue per unit sold which allows relatively inefficient firms to cover fixed costs and remain active,
thereby weakening selection. Fixed overhead costs, by contrast, shift profits
proportionally across firms and leave relative profitability unchanged, implying
that $\Gamma$ is invariant to $\phi$. That is, $\phi$ shifts the truncation point, but does not affect the shape of the Pareto tail above the truncation. 

Proposition~\ref{prop:Pareto_CS} shows how this selection logic translates into
aggregate outcomes, whilst accounting for channels outside $\Gamma$. First, the production-labour share $u$ depends only on the
returns to scale and markup, not on the level of fixed costs. Fixed costs
affect how many firms operate, but not how active firms allocate labour
internally. As $\nu$ rises, optimal firm size increases and a larger fraction of
aggregate labour is allocated to production rather than overheads. As $\mu$
rises, firms optimally scale down, increasing duplication of overhead labour and
reducing the production-labour share.

Second, the number of active firms adjusts along the extensive margin in a way
that mirrors these forces. Higher fixed costs and higher returns to scale both
reduce the equilibrium mass of firms, albeit for different reasons: fixed costs
mechanically discourage firms choosing to operate, while higher returns to scale raise optimal firm
size. Higher markups increase the number of firms by sustaining marginal
producers through higher per-unit revenues, generating an excess-entry
distortion.

Third, aggregate factor shares inherit these properties. The labour share is
invariant to fixed costs because adjustments in firm mass exactly offset changes
in overhead. When fixed costs increase, the number of firms fall, keeping the aggregate fixed cost share constant. The labour share falls with returns to scale, reflecting greater size diminishing the non-production (overhead) labour share more proportionally than increasing the production labour share. The labour share rises with markups, reflecting reduced
firm scale and greater labour overhead duplication. 

Lastly, the aggregate profit
share is independent of both fixed costs and returns to
scale. The independence of profits and total costs from $\phi$ holds in our general (distribution free) aggregate shares. Fixed costs only affect the choice of activity, but not input demands and factor shares. In the Pareto case, $\nu$ has no effect on the profit share (or total cost share) because as the labour share falls it is exactly offset by a rising capital share. By contrast, markups
through admitting smaller, low productivity, firms create a downward pressure on the profit share as factor shares expand through overheads.

Taken together, these results clarify the distinct economic roles of the three
key structural parameters. Fixed costs discipline entry but do not distort
within-firm allocation or income shares. Returns to scale reallocate resources
toward larger, more productive firms and strengthen selection $\Gamma$. Markups operate on
both margins simultaneously: they weaken selection, encourage excess entry,
reshape labour allocation, and compress aggregate profits as a share of output.
This decomposition highlights why the Pareto environment is particularly useful
for isolating the channels through which scale economies and market power affect
aggregate outcomes.

\subsection{Steady State Comparative Statics}

Under Pareto distributed technology the model exhibits a closed-form steady state. In particular we derive a closed-form solution for $\underbar{A}$ (Appendix Equation~\eqref{eq:Abar_SS_app}) from which we can infer the behaviour of other variables. Some comparative statics exhibit closed forms in others we use numerical illustrations based on the calibration from our quantitative exercise. In steady state, fixed costs and returns to scale both strengthen selection, but only the latter interact with markups through firms’ profitability and entry incentives. 

Changes in aggregate productivity occur through an extensive margin component $d \ln \Omega$ and an average productivity component $d \ln \hat{A}$:
\[d \ln TFP =  d \ln \Omega + d \ln \hat{A}\]

\subsubsection{The Effect of Entry Cost on Aggregate Productivity}

Although entry costs ($\kappa$) are not our focus, it is informative to discuss their role, particularly to distinguish them from fixed costs ($\phi$). Entry costs are output denominated and fixed costs are labour denominated. The entry cost affects aggregate productivity only through $\hat A$, leaving  $\Omega$ unchanged. A higher entry cost lowers $\hat A$ by reducing the productivity cutoff $\underline A$.\footnote{\textcite{BarseghyanDiCecio2011_JET} study this mechanism in a perfectly competitive economy and find that higher entry costs reduce aggregate TFP across countries.}

The mechanism follows from free entry. An increase in $\kappa$ raises the expected value required to enter, which is met by a lower cutoff productivity that increases the probability of being active ex post. In general equilibrium, higher entry costs reduce entry, lowering labour demand and wages. This reduces wage-denominated overhead costs, $w_t\phi$, enabling more active firms.

Importantly, entry costs and fixed operating costs have opposite effects on selection. Higher entry costs weaken selection by lowering the cutoff through wage adjustment, whereas higher operating costs $\phi$ directly raise the effective fixed cost ($w_t\phi$), increase the cutoff, and strengthen selection.

\subsubsection{The Effect of Fixed Costs on Aggregate Productivity}
Changes in fixed costs affect aggregate TFP as follows:
\[
\frac{d \ln TFP}{d \ln \phi}
=
-(1-\nu)
+
\frac{\nu(1-\alpha)}{\vartheta(1-\alpha \nu)-1}.
\]
The first term is $\frac{d \ln \Omega}{d \ln \phi} \gtreqless 0$ for $\nu \gtreqless 1$ and is independent of the productivity distribution. The second term captures the average productivity component and is strictly positive. Under Pareto, $\hat A=\Gamma \underline A$ with $\Gamma$ invariant to $\phi$, so higher fixed costs raise the productivity cutoff and strengthen selection. Notably, the competitive environment—summarised by the markup $\mu$—does not affect the impact of fixed costs on aggregate productivity.

\begin{figure}[H]
    \centering
    \singlespacing
    \caption{Effect of fixed costs on aggregate productivity by returns to scale}
    \begin{subfigure}[t]{0.48\linewidth}
        \centering
        \includegraphics[width=\linewidth]{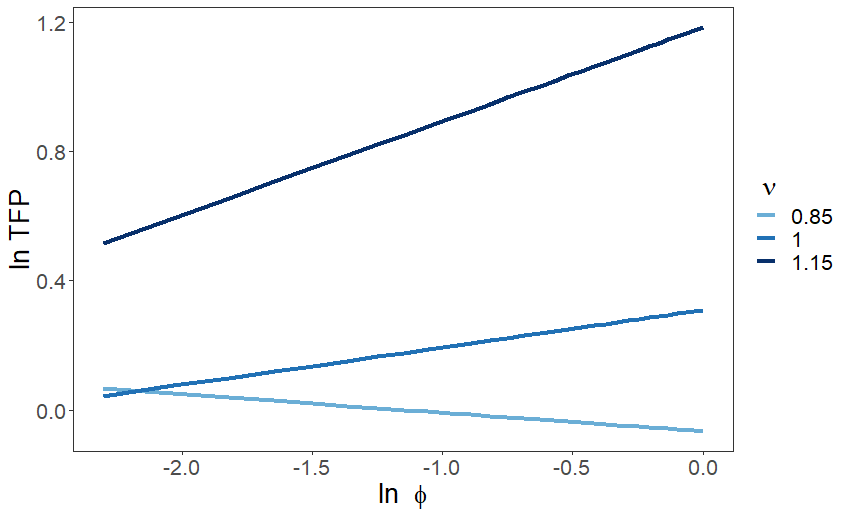}
        \caption{Aggregate measured productivity}
         \label{fig:lnTFP_lnphi_bynu}
    \end{subfigure}
    \hfill
    \begin{subfigure}[t]{0.48\linewidth}
        \centering
        \includegraphics[width=\linewidth]{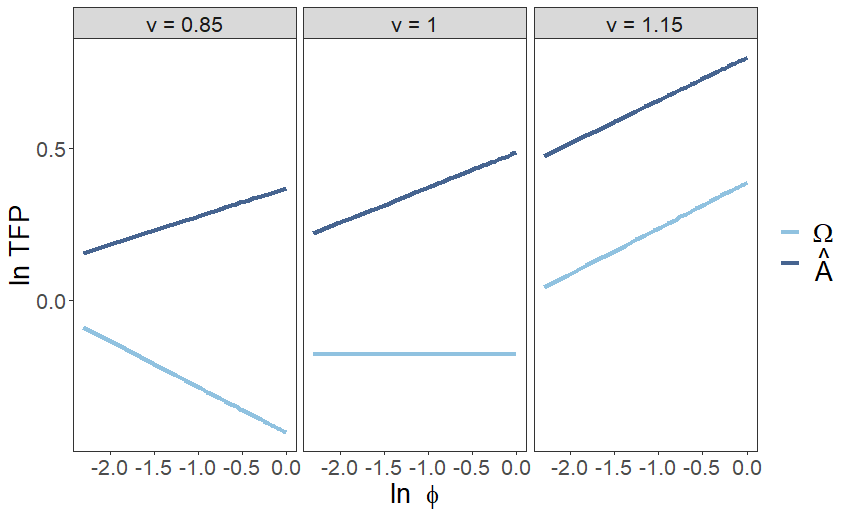}
        \caption{Decomposition into $\hat A$ and $\Omega$}
        \label{fig:lnTFP_decomp_lnphi_bynu}
    \end{subfigure}
    \caption*{\footnotesize ln TFP and its components for the calibrated model over a range of $\phi$ and $\nu$.}
\end{figure}

Figure \ref{fig:lnTFP_lnphi_bynu} plots steady-state measured TFP as a function of $\phi$ for three values of $\nu$ based on our benchmark calibration. 
In Figure \ref{fig:lnTFP_decomp_lnphi_bynu}, we decompose the three cases from Figure \ref{fig:lnTFP_lnphi_bynu}. Average productivity $\hat{A}_t$ always rises as the fixed cost increases, while the extensive margin component is determined by $\nu \gtreqless 1$.

\subsubsection{The Effect of Returns to Scale on Aggregate Productivity}

\begin{figure}[H]
    \centering
    \singlespacing
    \caption{Effect of returns to scale on aggregate productivity by markup}
    \begin{subfigure}[t]{0.48\linewidth}
        \centering
        \includegraphics[width=\linewidth]{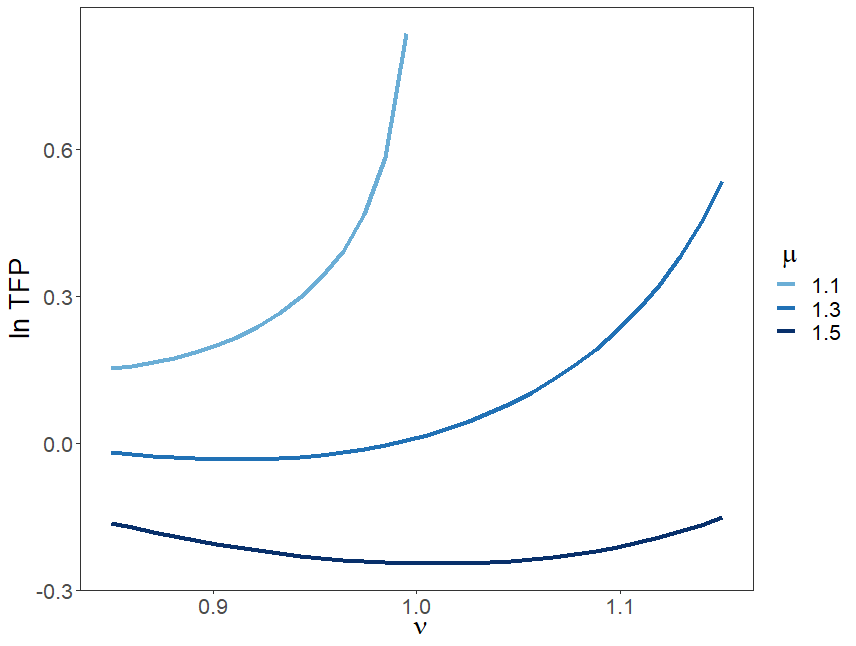}
        \caption{Aggregate measured productivity}
        \label{fig:TFP_nu_bymu}
    \end{subfigure}
    \hfill
    \begin{subfigure}[t]{0.48\linewidth}
        \centering
        \includegraphics[width=\linewidth]{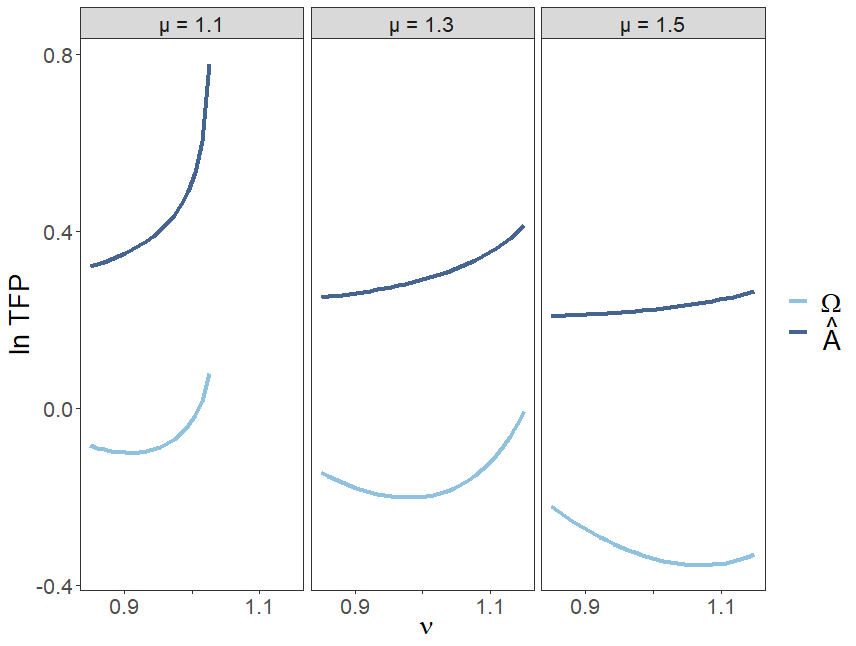}
        \caption{Decomposition into $\hat A$ and $\Omega$}
        \label{fig:TFP_nu_bymu_decomp}
    \end{subfigure}
    \caption*{\footnotesize ln TFP and its components for the calibrated model over a range of $\nu$ and $\mu$.}
\end{figure}

The steady-state solution for $\underline{A}$ is nonlinear in $\nu$, see Appendix equation~\eqref{eq:Abar_SS_app}. We illustrate the effect of returns to scale on aggregate productivity numerically. Figure~\ref{fig:TFP_nu_bymu} shows how aggregate productivity responds to $\nu$ for different values of the markup $\mu$. Aggregate productivity rises unambiguously in $\nu$ when markups are low, but this relationship weakens as markups increase: both the level and the slope of the response are declining in $\mu$.

Figure~\ref{fig:TFP_nu_bymu_decomp} decomposes this relationship into average productivity and extensive margin components. The muted passthrough of returns to scale to measured TFP at higher markups reflects both weaker selection (a flatter response of $\hat A$) and a weaker extensive margin effect. While higher $\nu$ strengthens selection by raising $\hat A$, this effect is increasingly dampened as market power rises.

\subsection{Empirical Relevance of Comparative Statics}

We document a set of stylised facts that are consistent with the qualitative comparative statics of the model. Using 2-digit SIC estimates of returns to scale, we relate cross-sector variation in $\nu$ to differences in the labour share, the number of active firms, and selection (productivity gap).

\begin{figure}[H]
    \centering
    \caption{Higher returns to scale ($\nu$) sectors have lower labour shares ($s_L$)}
    \includegraphics[width=0.7\linewidth]{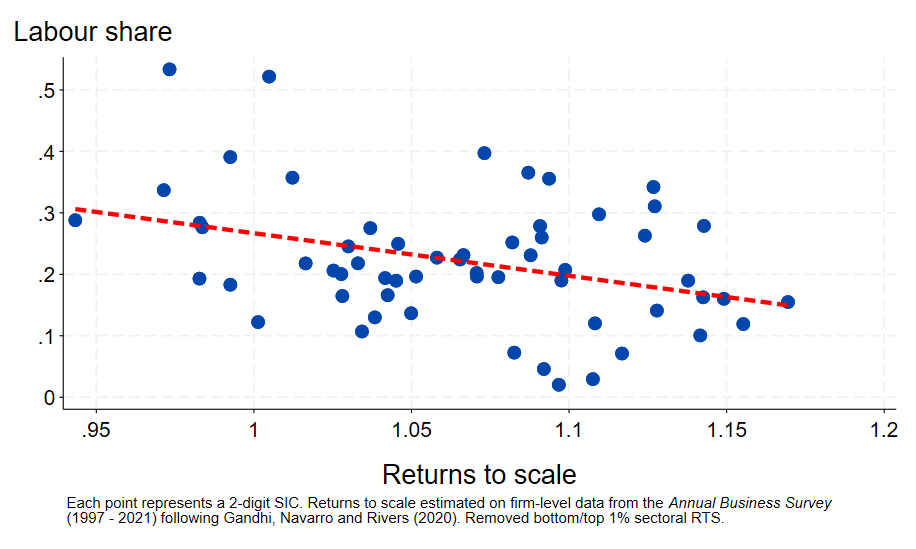}
    \label{fig:lab_share_rts_relationship}
\end{figure}

\begin{figure}[H]
    \centering
    \caption{Higher returns to scale ($\nu$) sectors have fewer active firms ($N$)}
    \includegraphics[width=0.7\textwidth]{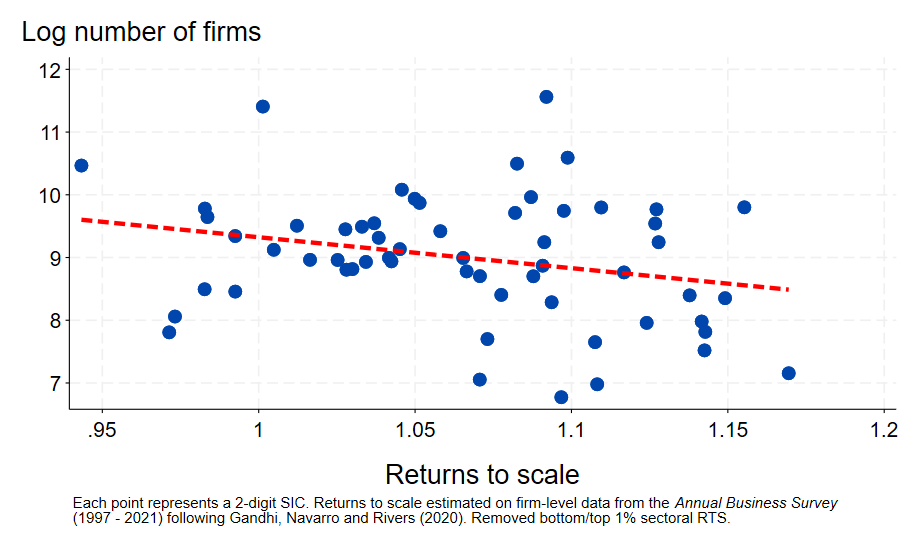}
    \label{fig:N_rts_relationship}
\end{figure}

\begin{figure}[H]
    \centering
    \caption{Higher returns to scale ($\nu$) sectors have a greater productivity gap ($\Gamma$)}
    \includegraphics[width=0.7\linewidth]{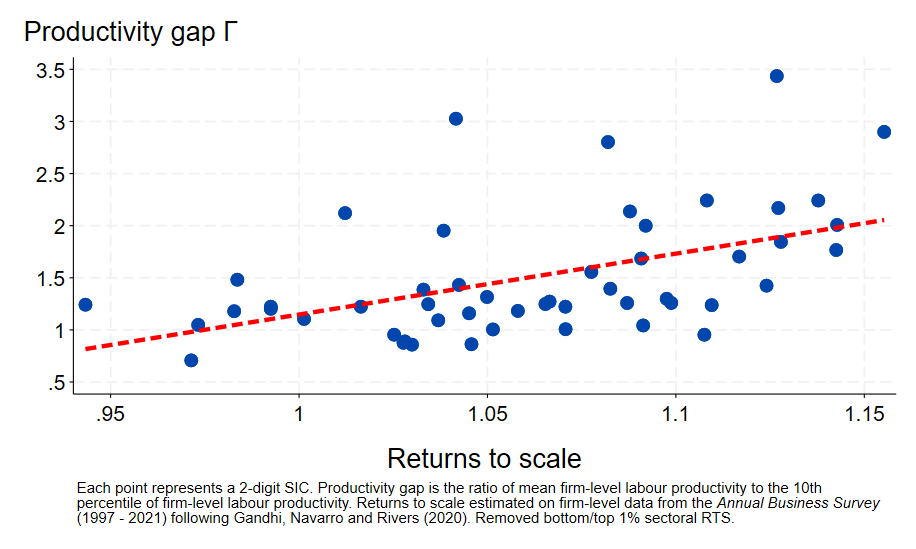}
    \label{fig:prod_gap_rts_relationship}
\caption*{\footnotesize The mean–p10 labour productivity ratio proxies firm selection, corresponding in the model to the ratio of average firm-level TFP to threshold TFP. More generally, it captures dispersion in firm quality.}
\end{figure}

The model predicts that higher returns to scale strengthen selection and reduce active firms, leading to less labour absorbed by fixed costs. As a result, the total labour share declines, but (under Pareto) the capital share rises proportionally, leading to no effect on the total cost share, or its residual: the profit share. Figures~\ref{fig:lab_share_rts_relationship}--\ref{fig:prod_gap_rts_relationship} are consistent with the qualitative comparative statics of the model. Sectors with higher estimated returns to scale exhibit systematically lower labour shares, fewer active firms, and larger productivity gaps (stronger selection). These cross-industry patterns provide stylised facts that align closely with the mechanisms highlighted by the model.

In summary, the qualitative equilibrium analysis yields three central conclusions. Selection matters for aggregates only through the production–overhead labour margin. Returns to scale strengthen selection, while markups systematically weaken it. Finally, fixed operating costs shape aggregate productivity independently of market power, whereas the productivity effects of returns to scale are constrained by markups.

\section{Endogenous Markups}

The baseline model assumes a constant markup~$\mu$, which isolates the role of scale economies in shaping firm size, selection, and aggregate productivity. To capture how rising market power interacts with these mechanisms, we analyse a reduced-form, partial equilibrium, extension in which the markup increases with productivity, $\mu=\mu(A)$ with $\mu'(A)>0$, as would be the case with a Kimball aggregator or oligpolistic competition. This provides the key intuition of endogenous markups without the complications of a full oligopolistic or non-homothetic preferences framework, where aggregation and entry decisions become complex.\footnote{For example, \textcite{DeLoeckerEeckhoutMongey2021_NBER, EdmondMidriganXu2023_JPE} develop sophisticated quantitative models with endogenous markups where entry is simplified by fixing the mass of potential entrants, whereas entry is a core channel in our model. Similarly, \textcite{OppParlourWalden2014_JET} have heterogeneous firms and oligopolistic competition, but rule-out entry. A notable complexity is that a potential entrant must internalise their own-effect on market shares, markups and profits, when deciding to enter.} In the quantitative exercise, we maintain the constant-markup structure but feed in year-specific empirical markups $\mu_t$ as exogenous wedges, allowing the model to evaluate scale-economy mechanisms under the actual competitive conditions observed in the data.

Adapting the reduced-form profit equation~\eqref{eq:rf_profit} from our constant markups framework to endogenous markups implies firm $\jmath$ earns profits
\begin{equation}
    \pi(\jmath)
    =
    \left(1 - \frac{\nu}{\mu(A(\jmath))}\right)
    R(\jmath)
    \;-\;
    w \phi,
\end{equation}
where we define revenue by
$R(\jmath) \equiv p(\jmath)y(\jmath)$.
The zero-profit condition $\pi(\underbar{A})=0$ implies the revenue threshold
\begin{equation} \label{eq:rev_threshold}
    R(\underbar{A})
    =
    \frac{w\phi\,\mu(\underbar{A})}{\mu(\underbar{A})-\nu}.
\end{equation}
This expression generalises the constant-markup case and makes clear that higher fixed costs ($\phi$) or stronger returns to scale ($\nu$) raise the required revenue to survive, while a higher cutoff markup $\mu(\underbar{A})$ lowers it.

Since revenue is strictly increasing in productivity, $R'(A)>0$, the cut-off $\underbar{A}$ is implicitly defined by
\begin{equation} \label{eq:cutoff_condition}
    H(\underbar{A};\nu,\phi)
    \equiv
    R(\underbar{A})
    -
    \frac{w\phi\,\mu(\underbar{A})}{\mu(\underbar{A})-\nu}
    = 0.
\end{equation}
Take the total derivative of $H(\underbar{A};\nu,\phi)=0$ with respect to $\nu$ and $\phi$ gives $ \mathrm{d}\underbar{A}/\mathrm{d}\nu = -H_\nu / H_A$ and $ \mathrm{d}\underbar{A}/\mathrm{d}\phi = -H_\phi / H_A$. The relevant derivatives of \eqref{eq:cutoff_condition} are as follows
\begin{equation}
    H_\nu
    =
    -\, w\phi\,
    \frac{\mu(\underbar{A})}{[\mu(\underbar{A})-\nu]^2}
    < 0, \qquad     H_\phi
    =
    - \frac{w\,\mu(\underbar{A})}{\mu(\underbar{A})-\nu}
    < 0, \qquad 
    H_A
    =
    R'(\underbar{A})
    +
    w\phi
    \frac{\nu\,\mu'(\underbar{A})}{[\mu(\underbar{A})-\nu]^2}
    > 0.
\end{equation}
Therefore, the comparative statics are
\begin{equation} \label{eq:dAdsigma}
    \frac{\mathrm{d}\underbar{A}}{\mathrm{d}\nu}
    =
    \frac{w\phi\,\mu(\underbar{A})}{[\mu(\underbar{A})-\nu]^2}
    \cdot \frac{1}{H_A}
    > 0 \qquad \frac{\mathrm{d}\underbar{A}}{\mathrm{d}\phi}
    =
    \frac{w\,\mu(\underbar{A})}{\mu(\underbar{A})-\nu}
    \cdot \frac{1}{H_A}
    > 0.
\end{equation}
Thus, stronger returns to scale or fixed costs tighten selection by raising the cut-off productivity. However, endogenous markups \emph{dampen} this effect. Under constant $\mu$, we would have $H_A = R'(\underbar{A})$; when $\mu'(\underbar{A})>0$, the additional positive term in $H_A$ increases the denominator in \eqref{eq:dAdsigma}, reducing the magnitude of the selection response.

Overall, endogenising markups introduces a countervailing force: scale economies push selection toward larger, more productive firms, but the rising markup schedule $\mu(A)$ cushions low-productivity firms by raising equilibrium profit as $A$ increases, thereby dampening the productivity gains.

\section{Quantitative Application} \label{sec:quantitative_analysis}

In Section \ref{sec:theoretical_analysis}, we examined the impact on aggregate productivity of the parameters of the production function that cause scale economies. We concluded that either higher fixed costs or higher returns to scale tend to increase aggregate productivity, though this relies on increasing returns $\nu>1$ to be unambiguous. We now analyse the quantitative plausibility of scale economies alongside stagnating productivity, which has occured in the US and UK in the first two decades of the 21st century. We find that changing returns to scale in variable inputs alongside rising markups explains the data well.

\subsection{Calibration}
Table \ref{tab:calibration_full} lists our calibrated parameters and how they are chosen in order to match observed empirical moments.
\begin{table}[H]
\centering
\caption{Model Fit to Calibration Targets}
\label{tab:calibration_full}
\begin{threeparttable}
\footnotesize
\begin{tabular}{llll}
\toprule
\textbf{Moment} & \textbf{Data} & \textbf{Value} & \textbf{Parameter} \\
\midrule
Real interest rate & 2.08\% (UK long-term real rate) & 0.96 & $\beta$ -- Discount rate \\
Depreciation rate & 8\% (ONS) & 0.08 & $\delta$ -- Depreciation rate \\
Capital share in rev. $\frac{\alpha \nu}{\mu}$ & 0.21 (ARDx) & 0.25 $\left(\frac{\nu}{\mu} \approx 0.85\right)$ & $\alpha$ -- Capital exponent \\
Returns to scale & 1.03--1.09 (ARDx) & 1.03--1.09 & $\nu$ -- Returns to scale \\
Markup & 1.21--1.36 (\textcite{CMA2022_report}) & 1.21--1.36 & $\mu$ -- Markup \\
Inactive firm share $J=1-\frac{N}{E}$ & 10\% (UK reactivation data) & 0.55 & $\phi$ -- Overhead cost \\
Workers per firm $L/N$ & 5.88--7.94 (Bus. Pop. Estimates) & 10 & $\vartheta$ -- Pareto tail \\
Ratio of fixed costs & $\frac{\kappa}{w\phi} < 1$ (industry evidence) & 0.275 & $\kappa$ -- Entry cost \\
\bottomrule
\end{tabular}
\begin{tablenotes}
\footnotesize
\item \textit{Notes:} Sources: ONS (Office for National Statistics); ARDx (Authors' calculations using ARDx); \textcite{CMA2022_report}; Business Population Estimates for firms-per-worker. Model moments come from steady-state equilibrium. The inactive firm share $J$ targets the fraction of firms that temporarily report zero production but reactive within two years. 
\end{tablenotes}
\end{threeparttable}
\end{table}

We set the discount factor $\beta$ to match the average real interest rate of 2.08 percent over the period. To do this, we use the equation for the steady-state interest rate $r = \frac{1}{\beta} + 1 - \delta$, where data on UK long-term government bonds and inflation are used to compute the real interest rate.\footnote{The FRED codes are: interest rates \href{https://fred.stlouisfed.org/series/IRLTLT01GBM156N}{IRLTLT01GBM156N} \parencite{OECD_FRED_IRLTLT01GBM156N} and inflation \href{https://fred.stlouisfed.org/series/FPCPITOTLZGGBR}{FPCPITOTLZGGBR} \parencite{WorldBank_FRED_FPCPITOTLZGGBR}.} The depreciation rate $\delta$ is determined by a weighted-average from ONS data. Our estimates of the returns to scale $\nu$ come from our estimates of the production function using the estimation of \textcite{GandhiNavarroRivers2020_JPE}.
Markup estimates are from \textcite{CMA2022_report}. They use a different dataset and estimation strategy. These markup estimates are consistent with other studies  that show rising markups over this time period \parencite{ONSBlack2022_WP,KarielSavagarHwang2022_WP}. Our results are not sensitive to these markup estimates. In the model $\frac{\alpha \nu}{\mu}$ is the capital share in revenue and $\frac{(1 - \alpha) \nu}{\mu}$ is the production labour share in revenue. Given our $\nu$ and $\mu$ estimates, we set $\alpha = 0.25$ to match a capital share of 21\%. 

The entry cost parameter $\kappa$ and the fixed cost parameter $\phi$ must satisfy restrictions such that $J_t \in (0,1)$. We choose $\phi$ to target the share of `inactive' firms $J_t$, to match the share of firms that do not produce but `re-activate' within a two year window. This is a standard approach used by the ONS and OECD to ensure accurate measures of firm deaths \parencite{ONS2025_BusinessDemography}. In the UK, the average share of `inactive' firms between 2016 and 2020 was 10\%, which is the only available time frame. We calibrate $\kappa$ by targeting a plausible entry-to-overhead cost ratio $\kappa / \phi w $. \textcite{BarseghyanDiCecio2011_JET} report a range of values from industry studies. In most industries, the ratio is less than one, so entry costs are less than overhead costs. The average they report is 0.82, which is also used by \textcite{AsturiasHurKehoeRuhl2023_AEJM}. Our experiments vary $\nu, \mu$ parameters, so the entry-to-overhead cost ratio will vary as we change these values, but the outcome always remains below 1.

To calibrate the Pareto shape $\vartheta$ we use the firm size distribution. Specifically, we use average firm size $L/N$ combined with our structural equation for aggregate overhead labour $1-u=\phi N/L = \left( \frac{\vartheta (\mu - \nu) - 1}{\vartheta (\mu - \alpha \nu) - 1} \right)$. We set the fixed parameters $\{\phi, \alpha\}$ as calibrated, time-varying parameters $\{\nu_t ,\mu_t\}$ as estimated, and $\{L_t, N_t\}$ timeseries from public data, detailed in the appendix. Following this approach, leads us to calibrate $\vartheta=10$. The economically-meaningful tail parameter in the model is the scaled Pareto tail $\vartheta^\ast\equiv\vartheta(\mu-\nu)$, which governs the firm size distribution. Given average $\mu_t-\nu_t$ then $\vartheta^\ast = 1.8$. Our calibration is similar to the US literature which targets the US firm size distribution in terms of employment. For example, the tail parameter determining US employment distribution is $1.15$ in \textcite{BarseghyanDiCecio2011_JET}, $1.06$ in \textcite{Luttmer2007_QJE}, $6.10$ in \textcite{AsturiasHurKehoeRuhl2023_AEJM}, and \textcite[p.743]{Hopenhayn2014_ARE} uses a 5 to 10 range for the raw Pareto tail.\footnote{In the appendix, we provide further calibration discussion.}

\subsection{Model Performance}

Table~\ref{tab:calibration_nontargeted} contains the non-targeted moments produced by the calibrated model. We consider the labour share, the profit share, the number of firms, the share of labour in production, and the capital-output ratio. 

\begin{table}[H]
\centering
\caption{Model Fit to Non-Targeted Moments}
\label{tab:calibration_nontargeted}
\begin{threeparttable}
\footnotesize
\begin{tabular}{lllllll}
\toprule
\textbf{Moment} & \multicolumn{3}{c}{\textbf{Data}} & \multicolumn{3}{c}{\textbf{Model}}  \\
& 2000 & 2019 & $\Delta$ (p.p. or \%) & 2000 & 2019 & $\Delta$ (p.p. or \%) \\
\midrule

Labour share in value added ($s_L$) 
  & 0.630 & 0.625 & -0.005 & 0.667 & 0.664 & -0.003  \\

Profit share ($s_\pi$) 
  & 0.125 & 0.108 & -0.017 & 0.082 & 0.073 & -0.009 \\

Production-labour share in total labour ($u$) 
  & 0.932 & 0.912 & -0.020 & 0.899 & 0.824 & -0.075 \\

Number of firms ($N$)
  & 1.11 & 1.41 & +26.9\% & 0.183 & 0.319 & +74.2\% \\

Capital--output ratio ($K/Y$) 
  & 1.95 & 1.96 & +0.43\% & 1.74 & 1.64 & -5.67\%  \\

\midrule
\multicolumn{7}{p{\linewidth}}{
\textit{Notes:} Labour share from \textcite{Caswell2024_WP}; profit share from  \href{https://www.ons.gov.uk/economy/nationalaccounts/uksectoraccounts/datasets/profitabilityofukcompaniesreferencetable}{ONS profitability of UK companies}; production-labour share is the share of labour not in administrative and support services (N) from the \href{https://www.ons.gov.uk/employmentandlabourmarket/peopleinwork/earningsandworkinghours/datasets/averageweeklyearningsbyindustryearn03}{ONS EARN03 employment weights}; number of firms (millions of employers) from \href{https://www.gov.uk/government/statistics/business-population-estimates-2022/business-population-estimates-for-the-uk-and-regions-2022-statistical-release-html}{UK Business Population Estimates}; capital-output ratio from \href{https://www.ons.gov.uk/economy/nationalaccounts/uksectoraccounts/datasets/capitalstocksconsumptionoffixedcapital}{ONS net capital stock} and \href{https://www.ons.gov.uk/economy/grossdomesticproductgdp/timeseries/abmi/pn2}{GDP}, chained volume measures. None of the moments in this table are used directly as calibration targets; they provide validation of the quantitative fit.} \\
\bottomrule
\end{tabular}
\end{threeparttable}
\end{table}

Table~\ref{tab:calibration_nontargeted} compares the data and the model in 2000 and 2019, where we calibrate returns to scale $\nu$ and the markup $\mu$ according to estimates from firm-level microdata. The model fits the data remarkably well given only two parameters are calibrated. The labour share is almost unchanged at 63\% in the data; it barely shifts from 67\% in the model. The profit share falls from 12.5\% to just below 11\% in the data; likewise it falls from around 8\% to 7\% in the model. The share of labour used in production (i.e. not in administrative or support roles) falls from 93\% to 91\% in the data. In the model the levels are close (90\% of labour is used in production in 2000) but the decline is 7.5 percentage points. The number of firms rises by substantially more in the model (75\%) than in the data (27\%). This is because it is entirely determined by the production share in labour ($N=\frac{1 - u}{\phi}$), which fell more sharply in the model than the data. Finally, the capital-output ratio sits just below 2 in the both years in the data; it is slightly lower in the model at 1.74 and falls by 5\%.
\subsection{Rising Returns to Scale Counterfactuals}
In our simulations, we focus on the ability of the model to generate measured TFP consistent with the data as we vary returns to scale and markups. We calibrate the parameter $\nu$ to our annual estimates from 2000 to 2019, while the parameter $\mu$ is set to annual estimates from \cite{CMA2022_report}. 

Figure~\ref{fig:calib_TFPindex_fixedmu} compares the trends in TFP in the data and our model. We present two data series. Data (PWT) is from the Penn World Tables as reported in our motivation (Figure~\ref{fig:TFP_growth}). This follows a traditional Solow Residual approach. Data (authors') is our model consistent measure of TFP based on equation~\eqref{eq:Y_CobbDouglas}, and estimated using ONS aggregate capital and hours data, combined with our calibration for $\alpha$ and $\nu_t$. The model and data trends show a rise in both series prior to the Financial Crisis, followed by a sharp decline in the data and a more gradual decrease in the model. The two counterfactual exercises act as bounds on the data and model, which highlights the opposing effects of returns to scale and markups. Importantly, the model is delivers measured TFP in between these bounds similar to both data series.

The counterfactual exercises are informative, highlighting the tug of war between productivity-enhancing returns to scale and productivity-degrading markups. Fixing the markup to its 2000 value highlights the significant impact of rising returns to scale on aggregate productivity. If market power had remained constant, higher returns to scale would have raised aggregate productivity by around 20\% between 2000 and 2019. However, when we incorporate the simultaneous increase in markups and returns to scale, our estimated productivity trend aligns more closely with observed data. Likewise, if returns to scale had not risen from 2000 onwards, but markups had increased as observed, aggregate productivity would have declined over 10\% over this period.
\begin{figure}[H]
\singlespacing
\centering
    \begin{tikzpicture}
\begin{axis}[
    width=0.95\linewidth,
    height=7cm,
    xlabel={Year},
    ylabel={Index (2000=100)},
    xmin=2000, xmax=2019,
    xtick={2000,2002,2004,2006,2008,2010,2012,2014,2016,2018},
    xticklabel style={/pgf/number format/1000 sep={}},
    ymajorgrids=true,
    grid style={draw=gray!25},
    legend columns=2,
    legend style={at={(0.02,0.02)}, anchor=south west},
]

\pgfplotstableread[col sep=comma]{./tikz_pgf/TFP_simulation.csv}\tfpdata

\addplot+[mark=none, thick, dashed, smooth]
table[x=year, y=model]{\tfpdata};
\addlegendentry{Model}

\addplot+[mark=none, thick, dotted, smooth]
table[x=year, y=model_fixmu]{\tfpdata};
\addlegendentry{Model (fixed $\mu$)}

\addplot+[mark=none, thick, dashdotted, smooth]
table[x=year, y=model_fixnu]{\tfpdata};
\addlegendentry{Model (fixed $\nu$)}

\addplot[smooth, thick, black, solid]
table[
    x=year,
    y=aggTFP_scaled,
    col sep=comma
]{\tfpdata};
\addlegendentry{Data (PWT)}

\addplot[smooth, thick, black, dotted]
table[
    x=year,
    y=TFP,
    col sep=comma
]{\tfpdata};
\addlegendentry{Data (authors')}
\end{axis}
\end{tikzpicture}
\caption{TFP index (2000=100): data vs model counterfactuals.}
    \label{fig:calib_TFPindex_fixedmu}
\caption*{\footnotesize We give the model estimates of $\mu$ and estimates of $\nu$ and solve in each year for steady-state to obtain the model-implied TFP. The TFP data series corresponds to Figure~\ref{fig:TFP_growth} \parencite{FRED_TFP_UK_RTFPNAGBA632NRUG}.}
\end{figure}

\subsubsection{Rising Overheads and Aggregate Productivity}

The joint rise in returns to scale, $\nu$, and markups, $\mu$, accounts well for aggregate productivity growth in the UK. At the same time, our empirical motivation in Figure \ref{figure:adshare_desc_stat} documents a sustained increase in administrative costs—measured either as a share of revenue or as labour-specific overheads—over the past two decades. This raises the question of whether rising overhead costs themselves can account for aggregate TFP dynamics in the UK.

Our comparative statics indicate that changes in $\nu$ generate richer aggregate implications than changes in the fixed cost parameter $\phi$. While $\phi$ affects productivity only through the extensive margin—by altering firms’ entry and production decisions—$\nu$ additionally reshapes internal production choices at the firm level. To supplement this insight, we extend the quantitative exercise by allowing $\phi$ to vary over time alongside $\nu$ and $\mu$. Rather than targeting the share of inactive firms $J$, as in the baseline calibration, we discipline $\phi$ using the equilibrium condition $\phi = \frac{1-u}{N/L}$,
which equates fixed overhead labour per firm to the ratio of aggregate non-production labour to the number of firms per worker.\footnote{Time-series data on firm reactivation rates in the UK—the moment used in our baseline calibration—are unavailable, whereas annual data on non-production labour shares and firms per worker are observed.}

Although the labour share in administrative and support services increases over time (Figure~\ref{figure:adshare_desc_stat}), this rise is more than offset by the rapid growth in the number of firms per worker, implying a decline in the calibrated fixed cost parameter $\phi$.\footnote{The labour share in administrative and support services rises from under 7\% in 2000 to 9\% in 2019 (Figure \ref{figure:adshare_desc_stat}), while the number of firms per worker increases from 0.126 to 0.179 (Table~\ref{tab:calibration_full}: $(L/N)^{-1}$), implying a 31\% decline in $\phi$.} Allowing for this time variation in $\phi$ leads to slightly lower aggregate TFP by 2019, but leaves the overall trend largely unchanged. This outcome is consistent with the model mechanism: lower fixed costs weaken selection, raise the number of active firms, and reduce average productivity.

Crucially, the decline in $\phi$ under this measure reflects firm proliferation rather than falling aggregate fixed costs.
The evidence therefore suggests that rising aggregate overheads is concentrated among larger firms in the upper tail of the size distribution, while the entry of small and young firms contributes relatively little. 

The key implication is that aggregate productivity dynamics are governed by how returns to scale and markups shape firm behavior, while rising aggregate overhead costs, concentrated among larger firms, play a secondary role in our framework.

\section{Conclusion}

This paper investigates the relationship between firm-level scale economies and aggregate productivity. We find evidence that returns to scale and fixed costs, which both determine scale economies, have increased since 1998 in the UK. To explore this relationship, we develop a theoretical framework linking firm-level fixed costs and returns to scale to aggregate productivity. We demonstrate that scale economies stemming from either fixed costs or returns to scale in variable inputs each have distinct implications for aggregate productivity due to their effects on firm selection and resource allocation.

Our model simulations reveal that while changing returns to scale offer a more plausible explanation for rising scale economies than rising fixed costs, these changes should have significantly boosted aggregate productivity.  This predicted growth is not reflected in the data. We resolve this discrepancy by highlighting evidence of increased markups within the UK over the same period, a factor that counteracts the productivity-enhancing effects of returns to scale. We show that if markups have the potential to rise endogenously as the number of firms fall, then it can offset the productivity gain from returns to scale.

In conclusion, our findings suggest that the combined effects of higher scale economies and increased market power help explain the stagnation of productivity growth seen during this period.

\newpage
\printbibliography

\newpage
\section*{Appendix}
\appendix

\section{Data}

\subsection{Fixed Cost Share Data}
We use the administration expenses share in turnover to proxy the fixed cost share for UK firms.
Figure \ref{figure:adshare_desc_stat} shows the median administration expenses share in turnover for UK firms from 2004 to 2023. 

\subsubsection*{Administrative Expenses}
In UK company accounts, ‘Administrative Expenses’ are defined as expenses an organization incurs that are not directly related to a specific function such as manufacturing, production, or sales. These expenses can include things like: rent, utilities, insurance, wages and benefits for administrative staff, depreciation on office furniture and equipment, professional fees (e.g., accounting and legal fees), and travel expenses. They are necessary for the day-to-day operation of a business, but they do not directly contribute to the generation of revenue. Expenses related to the generation of revenue fall under cost of goods sold (COGs). Administration expenses are typically reported on a company's income statement, below the cost of goods sold (COGS) line.

\subsubsection*{FAME data}
We use the Bureau van Dijk FAME dataset, a UK version of Orbis, to obtain firm financial information. The dataset records the annual financial statements of all incorporated companies in the UK. Over the entire period, there are 16,426,460 company entries. We restrict our analysis to companies that have at least one entry in administration expenses for any year between 2004 and 2023. The company does not need to be active today; it could have dissolved. This restriction reduces the number of companies to 680,763. The companies removed in this step have no administration expenses recorded over the sample period. This occurs because smaller companies can submit micro-entity accounts which do not include this information. Medium and large companies submit `full accounts' which do record this information. Due to download restrictions, we take a random sample of 250,000 companies, and we keep this same sample of firms every year. Since a firm only needs to have an administration expense in one year, there will be many blanks in any given year for any given company, either because it is inactive or because administration expenses were not recorded because it is a micro-entity. In the end, there are approximately 50,000 firms each year that have an entry in both administration expenses and turnover.

\subsection{Production Survey (ARD) Data}\label{app:data_background}

We use the Annual Respondents Database (ARD) or the time-series version known as ARDx. The ARD is based on the Annual Business Survey (ABS). The ABS is an annual survey of firms in the UK economy. It is a core ONS product used in the construction of national accounts. The ARD adds information from other business surveys to the ABS data.\footnote{Specifically, the ARD brings together the ABS and the Business Register and Employment Survey (BRES), and prior to 2009 it brought together the two parts of the Annual Business Inquiry (ABI).} 
Firms are legally obligated to respond to the survey. The survey forms a firm-level panel that covers all large firms and a representative sample of small firms by geography, size and sector. Large firms are surveyed annually, while small firms are surveyed for a fixed number of years.
The \href{https://beta.ukdataservice.ac.uk/datacatalogue/studies/study?id=7989}{ARDx Methodology} and \href{https://www.ons.gov.uk/businessindustryandtrade/business/businessservices/methodologies/annualbusinesssurveyabs}{ABS Methodology} provide more detail.
We use data covering 1997 - 2021.

\subsection{Capital Construction}

The Perpetual Inventory Method (PIM) allows the construction of firm-level capital stocks when such data are unavailable, but investment data is present. The method here follows \textcite{MartinCapitalStock} and \textcite{KarielSavagarHwang2022_WP}. The PIM is constructed using the following equation:
\[K_t=(1-\delta)K_{t-1} + I_t.\]
$K_t$ is the capital stock in period $t$, and $I_t$ is investment in period $t$. However, to use this method, we need $K_0$ -- the initial capital stock of a company, which is not in this survey. To construct this series, each firm's $K_0$ is a revenue-weighted share of the industry-level capital stock in the first year that firm appears in the panel. The capital stock is then constructed for all future years with the above equation, with the missing investment data interpolated. The depreciation rate is taken to be 18.195\%, which is a weighted average of the ONS depreciation rates for the three different capital categories: Building, Vehicles, Other. 

\subsection{Deflating} 
We convert firm gross output and value added into real values using the \href{https://www.ons.gov.uk/economy/inflationandpriceindices/datasets/experimentalindustrydeflatorsuknonseasonallyadjusted}{ONS industry deflators}. Material inputs are deflated with the \href{https://www.ons.gov.uk/economy/inflationandpriceindices/datasets/producerpriceindex}{ONS producer price inflation data}. The capital stock is deflated with the \href{https://www.ons.gov.uk/economy/grossdomesticproductgdp/timeseries/ybfu/ukea}{ONS gross fixed capital formation deflator}. 

\subsection{Cleaning} 
For the purpose of our production function estimation, we exclude sectors: Agriculture, Public Sector, Finance \& Insurance, Education, and Health. Standard Industrial Classification (SIC) 2007 codes: A, K, O, P, Q. These sectors were excluded from the survey after 2012. K,O,P were fully excluded and A,Q had various subsectors excluded. We set out rules for SIC re-coding to ensure compatibility pre- and post-2007, when the classification is changed. For SIC codes post-2007, we divide the number by 1000 to match with pre-2007 codes. 
To avoid outliers, which may represent recording errors in the surveys, we winsorize firms with the top and bottom 0.1\% of factor shares in revenue ($M/Y$, $K/Y$, $L/Y$) in each year. We also drop negative values of logged $Y, K, L, M$. Table \ref{tab:data_cleaning} contains number of firms at each stage of the data cleaning process, along with the final number of observations for estimation.
\begin{table}[H]
\centering
\caption{Data Cleaning: Firms Dropped}
\label{tab:data_cleaning}
    \begin{threeparttable} 
    \onehalfspacing
    \begin{tabular}{lc} 
        \toprule
        & \# Firms \\
        \midrule
        All ARD firm-year obs & 1,152,014 \\
        Drop sectors A,K,O,P,Q & 1,071,014 \\
        Drop outlier/missing factor shares & 988,960 \\
        Take logs of regression variables & 812,941 \\
        \bottomrule
    \end{tabular} 
        \begin{tablenotes}
        \item \textit{\small{}}
        \end{tablenotes}
    \end{threeparttable}
\end{table}

\subsection{Summary Statistics}

Table \ref{tab:desc_stats_reg_variables_full_sample} presents aggregate descriptive statistics of the variables used in our regression analysis. 

\begin{table}[H]
\centering
\singlespacing
\caption{Descriptive Statistics of Regression Variables}
\label{tab:desc_stats_reg_variables_full_sample}
\begin{tabular}{@{}llllll@{}}
\toprule
 & Mean & p10 & p50 & p90 & $N$ \\ \midrule
Revenue    &  39,579   & 51    & 1,244 & 41,815    & 812,941 \\
Labour     &  243.4    & 2     & 25    & 367       & 812,941 \\
Capital    &  11,735   & 25.8  & 342.6 & 10,259    & 812,941 \\
Materials  &  30,140   & 16     & 625  & 26,796    & 812,941 \\
Labour Share    & 0.25 & 0.00  & 0.21  & 0.52      & 812,941 \\
Capital Share   & 0.32 & 0.05  & 0.25  & 0.72      & 812,941 \\
Materials Share & 0.51 & 0.09  & 0.54  & 0.86      & 812,941 \\
\bottomrule
\end{tabular}
\end{table}

\subsection{Returns to Scale Estimates} \label{appendix_RTS}

Table \ref{tab:rts_gnr_2d_1}-\ref{tab:rts_gnr_2d_2} presents production function estimates across 2-digit SICs, averaged over 1997 - 2021, following the approach of \textcite{GandhiNavarroRivers2020_JPE}. We present the underlying coefficients on labour, capital, and materials, which sum to returns to scale. The estimates of returns to scale range from 0.91 to 1.99. Of the 62 industries for which we obtained results, 45 of them obtain returns to scale between 0.90 and 1.10, and 24 of them are between 0.95 and 1.05.

\begin{table}[H]
\centering
\singlespacing
\caption{Production function estimates by 2-digit SIC}
\label{tab:rts_gnr_2d_1}
    \begin{threeparttable}
    \singlespacing
    \begin{tabular}{lccccc}
        \toprule
        2-digit SIC & $\beta_l$ & $\beta_k$ & $\beta_m$ & RTS & $N$ \\
        \midrule
        10 & 0.288 (0.011) & 0.152 (0.005) & 0.610 (0.009) & 1.050 & 20,674 \\
        11 & 0.482 (0.051) & 0.141 (0.041) & 0.518 (0.005) & 1.142 & 2,923 \\
        12 & 0.962 (0.760) & 0.813 (0.699) & 0.212 (0.049) & 1.986 & 106 \\
        13 & 0.372 (0.017) & -0.046 (0.016) & 0.691 (0.007) & 1.016 & 7,811 \\
        14 & 0.347 (0.026) & -0.018 (0.026) & 0.664 (0.012) & 0.993 & 4,709 \\
        15 & 0.480 (0.057) & 0.064 (0.033) & 0.527 (0.018) & 1.071 & 1,158 \\
        16 & 0.448 (0.022) & 0.051 (0.023) & 0.572 (0.008) & 1.071 & 6,021 \\
        17 & 0.312 (0.014) & 0.070 (0.013) & 0.646 (0.005) & 1.028 & 6,674 \\
        18 & 0.554 (0.009) & 0.047 (0.012) & 0.490 (0.006) & 1.091 & 10,351 \\
        19 & 0.354 (0.080) & 0.033 (0.068) & 0.710 (0.012) & 1.097 & 874 \\
        20 & 0.358 (0.008) & 0.078 (0.024) & 0.603 (0.009) & 1.038 & 11,102 \\
        21 & 0.625 (0.013) & 0.025 (0.038) & 0.493 (0.021) & 1.143 & 1,844 \\
        22 & 0.423 (0.010) & 0.001 (0.008) & 0.604 (0.004) & 1.028 & 12,705 \\
        23 & 0.411 (0.028) & 0.106 (0.023) & 0.528 (0.001) & 1.045 & 9,266 \\
        24 & 0.450 (0.011) & 0.020 (0.012) & 0.572 (0.004) & 1.042 & 7,621 \\
        25 & 0.517 (0.002) & 0.054 (0.003) & 0.474 (0.003) & 1.046 & 23,891 \\
        26 & 0.556 (0.028) & -0.018 (0.020) & 0.520 (0.005) & 1.058 & 12,331 \\
        27 & 0.483 (0.022) & 0.006 (0.010) & 0.536 (0.009) & 1.025 & 7,796 \\
        28 & 0.550 (0.010) & 0.018 (0.013) & 0.484 (0.008) & 1.051 & 19,362 \\
        29 & 0.411 (0.012) & 0.003 (0.008) & 0.620 (0.003) & 1.034 & 7,557 \\
        30 & 0.539 (0.006) & 0.006 (0.004) & 0.533 (0.006) & 1.078 & 4,467 \\
        31 & 0.414 (0.018) & 0.023 (0.028) & 0.592 (0.005) & 1.030 & 6,738 \\
        32 & 0.504 (0.026) & 0.097 (0.012) & 0.464 (0.006) & 1.065 & 8,050 \\
        33 & 0.572 (0.007) & 0.008 (0.012) & 0.486 (0.002) & 1.066 & 6,494 \\
        41 & 0.679 (0.005) & 0.126 (0.010) & 0.350 (0.012) & 1.155 & 18,048 \\
        42 & 0.525 (0.013) & 0.088 (0.006) & 0.485 (0.004) & 1.098 & 17,061 \\
        43 & 0.628 (0.021) & 0.055 (0.005) & 0.416 (0.004) & 1.099 & 39,817 \\
        45 & 0.334 (0.028) & 0.042 (0.008) & 0.706 (0.006) & 1.083 & 36,178 \\
        46 & 0.401 (0.016) & 0.092 (0.007) & 0.600 (0.004) & 1.092 & 104,991 \\
        47 & 0.339 (0.013) & -0.025 (0.005) & 0.687 (0.003) & 1.001 & 89,947 \\
        49 & 0.498 (0.029) & 0.068 (0.033) & 0.417 (0.008) & 0.983 & 17,686 \\
        \bottomrule
    \end{tabular}
        \begin{tablenotes}
        \item \textit{\small{Standard errors are reported in parentheses next to coefficients. RTS is returns to scale. $N$ is the number of firms.}}
        \end{tablenotes}
    \end{threeparttable}
\end{table}

\begin{table}[H]
\centering
\singlespacing
\caption{Production function estimates by 2-digit SIC}
\label{tab:rts_gnr_2d_2}
    \begin{threeparttable}
    \singlespacing
    \begin{tabular}{lccccc}
        \toprule
        2-digit SIC & $\beta_l$ & $\beta_k$ & $\beta_m$ & RTS & $N$ \\
        \midrule
        51 & 0.737 (0.062) & 0.090 (0.098) & 0.343 (0.056) & 1.169 & 1,281 \\
        52 & 0.523 (0.009) & 0.147 (0.013) & 0.363 (0.005) & 1.033 & 13,244 \\
        53 & 0.631 (0.020) & 0.036 (0.005) & 0.406 (0.002) & 1.073 & 2,209 \\
        55 & 0.457 (0.014) & 0.184 (0.011) & 0.395 (0.001) & 1.037 & 13,996 \\
        56 & 0.329 (0.006) & 0.151 (0.006) & 0.464 (0.004) & 0.943 & 35,117 \\
        58 & 0.637 (0.029) & 0.089 (0.027) & 0.364 (0.003) & 1.091 & 7,113 \\
        59 & 0.944 (0.069) & -0.082 (0.041) & 0.276 (0.006) & 1.138 & 4,434 \\
        60 & 0.455 (0.110) & 0.243 (0.034) & 0.411 (0.030) & 1.108 & 1,074 \\
        61 & 0.365 (0.051) & 0.023 (0.038) & 0.760 (0.017) & 1.149 & 4,239 \\
        62 & 0.764 (0.012) & 0.084 (0.008) & 0.280 (0.008) & 1.127 & 17,468 \\
        63 & 0.729 (0.037) & 0.157 (0.048) & 0.257 (0.009) & 1.143 & 2,478 \\
        69 & 0.715 (0.011) & 0.189 (0.005) & 0.183 (0.002) & 1.087 & 21,226 \\
        70 & 0.712 (0.015) & 0.200 (0.012) & 0.215 (0.006) & 1.127 & 13,935 \\
        71 & 0.717 (0.023) & 0.114 (0.008) & 0.279 (0.003) & 1.110 & 17,995 \\
        72 & 0.717 (0.017) & 0.043 (0.042) & 0.334 (0.014) & 1.094 & 3,972 \\
        73 & 0.462 (0.008) & 0.254 (0.007) & 0.325 (0.004) & 1.042 & 8,062 \\
        74 & 0.674 (0.031) & 0.118 (0.014) & 0.296 (0.004) & 1.088 & 6,011 \\
        75 & 0.563 (0.023) & 0.103 (0.021) & 0.306 (0.004) & 0.971 & 2,459 \\
        77 & 0.595 (0.052) & 0.332 (0.024) & 0.201 (0.029) & 1.128 & 10,344 \\
        78 & 0.608 (0.015) & 0.168 (0.018) & 0.128 (0.003) & 0.905 & 15,778 \\
        79 & 0.548 (0.012) & 0.121 (0.016) & 0.447 (0.009) & 1.117 & 6,393 \\
        80 & 0.639 (0.027) & 0.157 (0.031) & 0.178 (0.007) & 0.973 & 3,164 \\
        81 & 0.544 (0.003) & 0.224 (0.003) & 0.225 (0.003) & 0.993 & 11,412 \\
        82 & 0.663 (0.022) & 0.149 (0.011) & 0.270 (0.005) & 1.082 & 16,509 \\
        90 & 0.542 (0.040) & 0.118 (0.012) & 0.323 (0.022) & 0.983 & 4,888 \\
        91 & 0.428 (0.049) & 0.190 (0.081) & 0.626 (0.042) & 1.244 & 3,310 \\
        92 & 0.636 (0.081) & 0.120 (0.068) & 0.352 (0.019) & 1.108 & 2,103 \\
        93 & 0.398 (0.014) & 0.294 (0.004) & 0.320 (0.001) & 1.012 & 13,459 \\
        94 & 0.414 (0.056) & 0.186 (0.059) & 0.405 (0.019) & 1.005 & 9,171 \\
        95 & 0.703 (0.033) & 0.102 (0.023) & 0.319 (0.009) & 1.124 & 2,859 \\
        96 & 0.590 (0.026) & 0.149 (0.021) & 0.245 (0.005) & 0.984 & 15,452 \\
        \bottomrule
    \end{tabular}
        \begin{tablenotes}
        \item \textit{\small{Standard errors are reported in parentheses next to coefficients. RTS is returns to scale. $N$ is the number of firms.}}
        \end{tablenotes}
    \end{threeparttable}
\end{table}

\section{Equilibrium Conditions with Pareto Distribution}
We parameterise productivity using the quantile function. Let $\jmath$ be uniformly distributed on $[0,1)$, and define
\begin{equation}
A(\jmath) = \frac{h}{(1-\jmath)^{1/\vartheta}},
\end{equation}
where $\vartheta>1$ is the Pareto shape parameter and $h$ is the scale parameter. We normalise $h=1$, so $A(0)=1$ is the minimum productivity draw. A lower value of $\vartheta$ corresponds to a thicker (heavier) tail and therefore a greater mass of high-productivity firms; conversely, $\vartheta \to \infty$ collapses the distribution toward its lower bound. To ensure the Pareto distribution is well behaved after scaling $A(\jmath)$ by $1/(\mu-\nu)$, we impose two restrictions which limit the thickness of Pareto tails.
\begin{assumption}\label{ass:vartheta}
The Pareto shape parameter satisfies
\begin{equation}
    \frac{1}{\vartheta} < \min \left\{ \mu - \nu, 1 - \alpha \nu \right\}.
\end{equation}
\end{assumption}
The restriction $\vartheta(\mu-\nu) > 1$ guarantees that the expectation $\mathbb{E}[A(\jmath)^{1/(\mu-\nu)}]$ is finite, analagous to the standard Pareto requirement $\vartheta>1$ before scaling. The condition $1 - \vartheta(1-\alpha\nu) < 0$ ensures that aggregate production remains concave in aggregate capital, which in turn guarantees that the rental rate of capital is decreasing in the aggregate capital stock.

Under Pareto, the average technology is a linear function of the cutoff productivity:
\begin{equation} \label{eq:a_power_mean_pareto_revised}
\hat{A}_t =
\Gamma
\underline{A}_t
\end{equation}
The gap between average productivity and threshold productivity is a constant:
\[\Gamma = \left( \frac{\vartheta(\mu-\nu)}{\vartheta(\mu-\nu) - 1} \right)^{\mu-\nu}.
\]
The production–labour to total labour share and fixed-cost to total labour share are constant:
\begin{align*}
u &= 
\left(1 + \frac{\vartheta (\mu - \nu) -1}{\nu \vartheta (1-\alpha)} \right)^{-1}, 
\qquad
1-u = \frac{\vartheta(\mu - \nu)-1}{\vartheta(\mu - \alpha\nu)-1},
\end{align*}
Therefore, the number of firms is also constant
\[N= \frac{1-u}{\phi}.\]
Aggregate productivity is
\begin{equation}
TFP_t = \Omega\, \hat A_t \label{eq:pareto_system_TFP}
\end{equation}
where
\[\Omega = \left(\frac{1-u}{\phi}\right)^{1-\nu} u^{(1-\alpha)\nu},
\qquad \text{and} \qquad 
\hat A_t = \Gamma \,\underline{A}_t .\]
Free entry is:
\begin{equation}
    w_t = \frac{\kappa}{\phi}\big[\vartheta(\mu - \nu) - 1\big] \underline{A}_t^{\vartheta}.
\end{equation}
The remaining equilibrium conditions, which are unaffected by the distribution, are
\begin{align}
Y_t - C_t &= K_{t+1} - (1-\delta)K_t, \label{eq:Kdyn_app}\\
\left(\frac{C_{t+1}}{C_t}\right)^{\sigma} &= \beta \left[r_{t+1} + 1-\delta\right], \label{eq:Cdyn_app}\\
Y_t &= TFP_t\, K_t^{\alpha \nu}, \\
r_t &= \frac{\nu}{\mu}\alpha \frac{Y_t}{K_t}, \\
w_t &= \frac{\nu}{\mu}(1-\alpha)\frac{Y_t}{u}. \label{eq:pareto_system_w}
\end{align}
Thus the model consists of seven conditions (\eqref{eq:pareto_system_TFP} - \eqref{eq:pareto_system_w}) in seven unknowns:
\[
\{C_t,K_t,Y_t,r_t,w_t,TFP_t,\underline{A}_t\},
\qquad 
u \ \text{constant}.
\]

\subsection*{Reduction to a Two-Dimensional Dynamic System}

Equating wages from the factor-market condition and free entry, and substituting
\[
Y_t = TFP_t K_t^{\alpha\nu} = \Omega \Gamma \underline{A}_t K_t^{\alpha\nu},
\]
we obtain
\[
(1-\alpha)\frac{\nu}{\mu}\frac{\Omega \Gamma}{u}\,\underline{A}_t K_t^{\alpha\nu}
=
\frac{\kappa}{\phi}\big[\vartheta(\mu - \nu) -1\big] \underline{A}_t^\vartheta .
\]
For given $K_t$, the threshold $\underline{A}_t$ must adjust to clear the labour market.  
Because $\vartheta>1$, the free-entry wage is more sensitive to $\underline{A}_t$ than the factor-market wage. Hence higher capital raises wages and tightens selection: only firms with higher $A$ survive.

Solving for $\underline{A}_t$ gives
\begin{equation}\label{eq:Abar_sol_app}
\underline{A}_t 
= 
\Psi\, K_t^{\frac{\alpha\nu}{\vartheta-1}},
\qquad
\Psi 
\equiv 
\left[
\frac{\phi}{\kappa[\vartheta(\mu - \nu) -1]}
(1-\alpha)\frac{\nu}{\mu}\frac{\Omega \Gamma}{u}
\right]^{\!\frac{1}{\vartheta-1}}.
\end{equation}

Substituting \eqref{eq:Abar_sol_app}, all remaining equilibrium variables become functions of $K_t$:
\begin{align}
TFP_t &= \Omega \Gamma \Psi\, K_t^{\frac{\alpha\nu}{\vartheta-1}}, \label{eq:TFP_app}\\
w_t &= \frac{\kappa}{\phi}\big[\vartheta(\mu - \nu) -1\big] \Psi^\vartheta K_t^{\frac{\alpha \nu \vartheta}{\vartheta -1}}, \label{eq:w_app}\\
r_t &= \alpha\frac{\nu}{\mu}\Omega\Gamma\Psi\, K_t^{\frac{\alpha\nu\vartheta}{\vartheta-1}-1}, \label{eq:r_app}\\
Y_t &= \Omega \Gamma \Psi\, K_t^{\frac{\alpha\nu\vartheta}{\vartheta -1}}. \label{eq:Y_app}
\end{align}

Substituting into \eqref{eq:Kdyn_app}–\eqref{eq:Cdyn_app} yields the two-dimensional reduced system:
\begin{align}
\Omega \Gamma \Psi K_t^{\frac{\alpha\nu\vartheta}{\vartheta-1}} - C_t
&= K_{t+1} - (1-\delta)K_t, \label{eq:K_reduced_app}
\\
\left(\frac{C_{t+1}}{C_t}\right)^\sigma
&= 
\beta\left[
\alpha\frac{\nu}{\mu}\Omega\Gamma\Psi\,K_{t+1}^{\frac{\alpha\nu\vartheta}{\vartheta-1}-1}
+ 1-\delta
\right]. \label{eq:C_reduced_app}
\end{align}

\subsection*{Capital Elasticity}

The elasticity of the rental rate with respect to capital is
\[
\frac{d\ln r_t}{d\ln K_t}
=
\frac{1 - \vartheta(1-\alpha\nu)}{\vartheta -1}.
\]
Thus $r_t$ is decreasing in $K_t$ when 
\[
1 - \vartheta(1-\alpha\nu) < 0.
\]
This ensures aggregate production is concave in capital.  

\subsection{Steady State Under Pareto}

In steady state $K_{t+1}=K$ and $C_{t+1}=C$, and the Euler equation implies
\[
r = \frac{1}{\beta} - (1-\delta).
\]
Using \eqref{eq:r_app}, steady-state capital is
\[
K =
\left[
\frac{\alpha\nu\,\Omega\Gamma\Psi}{\mu r}
\right]^{\!\frac{\vartheta -1}{\vartheta(1-\alpha\nu)-1}},
\]
and steady-state consumption follows from the resource constraint:
\[
C = K\left(\frac{\mu r}{\alpha\nu} - \delta\right).
\]
Substituting back into the reduced system yields all other steady-state variables.  
The selection threshold is
\begin{align}\label{eq:Abar_SS_app}
\underline{A}
= 
\left[
\nu^\nu 
\frac{1}{\mu}
\left(\frac{\alpha}{r}\right)^{\alpha\nu}
\big(\phi(1-\alpha)\big)^{\nu(1-\alpha)}
\vartheta^{\mu-1}(\mu-\nu)^{\mu-\nu}
\frac{1}{\kappa^{1-\alpha\nu}}
\frac{1}{\big[\vartheta(\mu-\nu)-1\big]^{\mu-\alpha\nu}}
\right]^{\!\frac{1}{\vartheta(1-\alpha\nu)-1}}.
\end{align}

\subsubsection{Entry Cost Restriction}

Since productivity draws satisfy $\underline{A}\geq 1$, admissible parameter values must satisfy
\[
1 \le 
\left[
\nu^\nu 
\frac{1}{\mu}
\left(\frac{\alpha}{r}\right)^{\alpha\nu}
\big(\phi(1-\alpha)\big)^{\nu(1-\alpha)}
\vartheta^{\mu-1}(\mu-\nu)^{\mu-\nu}
\frac{1}{\kappa^{1-\alpha\nu}}
\frac{1}{\big[\vartheta(\mu-\nu)-1\big]^{\mu-\alpha\nu}}
\right]^{\!\frac{1}{\vartheta(1-\alpha\nu)-1}}.
\]
Equivalently, the entry cost must not exceed
\[
\kappa 
\le
\left[
\nu^\nu 
\frac{1}{\mu}
\left(\frac{\alpha}{r}\right)^{\alpha\nu}
\big(\phi(1-\alpha)\big)^{\nu(1-\alpha)}
\vartheta^{\mu-1}(\mu-\nu)^{\mu-\nu}
\frac{1}{\big[\vartheta(\mu-\nu)-1\big]^{\mu-\alpha\nu}}
\right]^{\!\frac{1}{1-\alpha\nu}}.
\]
If the restriction holds with equality, then $\underline{A}=1$ (so $J=0$), implying no selection and $N=E$. The upper bound on the entry cost $\kappa$ is the global analogue of the firm-level participation constraint $v(\jmath)=\max\{\pi(\jmath),0\}$. If $\kappa$ exceeds this bound, even the maximum attainable expected value—obtained when all entrants operate—falls short of the entry cost. In this case, free entry cannot be satisfied and equilibrium entry collapses.

\section{Comparative Statics Proofs}
\label{app:Pareto_CS_proofs}

\begin{proof}[Proof of Lemma~\ref{lem:Gamma_derivatives} (Pareto firm selection behaviour)]\label{pr}
Let $x \equiv \mu-\nu>0$ and note that by assumption $\vartheta x>1$. Define
\[
\Gamma(x)
=
\left( \frac{\vartheta x}{\vartheta x - 1} \right)^{x}= q(x)^{x},
\qquad
q(x) \equiv \frac{\vartheta x}{\vartheta x - 1} > 1.
\]
Taking logs followed by the derivative gives
\[
\frac{d}{dx}\ln \Gamma(x) = \ln q(x) + x \frac{q'(x)}{q(x)} = \ln q(x) + 1 - q(x).
\]
Taking the derivative of the left-hand side and rearranging yields:
\[
\frac{d\Gamma}{dx}
= \Gamma(x)\big(\ln q(x) + 1 - q(x)\big) < 0.
\]
The inequality follows from the standard log inequality, that for $q>1$ then $\ln q < q-1$. Finally, using $x=\mu-\nu$,
\[
\frac{\partial\Gamma}{\partial\nu}
=\frac{d\Gamma}{dx}(-1)>0,
\qquad
\frac{\partial\Gamma}{\partial\mu}
=\frac{d\Gamma}{dx}<0.
\]
\end{proof}

\begin{proof}[Proof of Proposition~\ref{prop:Pareto_CS} (Pareto Comparative Statics)]
Under Pareto technology, the production-labour share, number of firms, labour share, and profit share are:
\[
u=\left(1+\frac{\vartheta(\mu-\nu)-1}{\nu\vartheta(1-\alpha)}\right)^{-1},\qquad
N=\frac{1-u}{\phi},\qquad
s_L=\frac{1}{\mu}\left(\mu-\alpha\nu-\frac{1}{\vartheta}\right) ,\qquad
s_\pi=\frac{1}{\vartheta \mu}.
\]

\paragraph{(i) Comparative statics of $u$.}
Since $u$ does not depend on $\phi$,  
\[
\frac{d u}{d\phi}=0.
\]
Differentiation yields
\[
\frac{d u}{d\nu}
    =\frac{(1-\alpha)\vartheta(\vartheta\mu-1)}{[\vartheta(\mu-\alpha\nu)-1]^2}>0,
\qquad
\frac{d u}{d\mu}
    =-\frac{(1-\alpha)\vartheta^2\nu}{[\vartheta(\mu-\alpha\nu)-1]^2}<0,
\]
using $\vartheta(\mu-\nu)>1$ and $\alpha\in(0,1)$.

\paragraph{(ii) Comparative statics of $N$.}
\[
\frac{dN}{d\phi}=-\frac{N}{\phi}<0,\qquad
\frac{dN}{d\nu}=-\frac{1}{\phi}\frac{du}{d\nu}<0,\qquad
\frac{dN}{d\mu}=-\frac{1}{\phi}\frac{du}{d\mu}>0.
\]

\paragraph{(iii) Comparative statics of the labour share.}
Since $s_L$ does not depend on $\phi$,
\[
\frac{d s_L}{d\phi}=0.
\]
Differentiation gives
\[
\frac{d s_L}{d\nu}=-\frac{\alpha}{\mu}<0,
\qquad
\frac{d s_L}{d\mu}=\frac{1}{\mu}(1-s_L)>0,
\]
because $\alpha,\mu>0$ and $s_L<1$.

\paragraph{(iv) Comparative statics of the profit share.}
Since $s_\pi$ does not depend on $\phi$ or $\nu$,
\[
\frac{d s_\pi}{d\phi}=0, \qquad \frac{d s_\pi}{d\nu}=0.
\]
Differentiation gives
\[
\frac{d s_\pi}{d\mu}=-\frac{1}{\vartheta \mu^2}<0.
\]
\end{proof}

\section{Calibration Details}

\subsection{RTS to markup ratio}
The ratio $\nu/\mu$ is the elasticity of firm revenue to variable inputs; equivalently, it is the variable-cost share in revenue for the firm or the variable-cost share in total output for the aggregate economy. The remainder $1-\nu/\mu$ is the profit share plus the fixed cost share. The ratio $\nu/\mu$ is typically set to 0.85 in US studies \parencite{RestucciaRogerson2008_RED, BarseghyanDiCecio2011_JET, Hopenhayn2014_ARE} with perfect competition $\mu=1$. Our estimates for $\nu$ divided by our calibrated markup $\mu$ yield a ratio from 0.80 to 0.89 between 2000 and 2019.

\subsection{Pareto Tail Index}

We obtain active firms $N_t$ from business population estimates \parencite{BEIS2022_BusPopEst} and total employment $L_t$ from the ONS \parencite{ONS2025_MGRZ}. From 2000 to 2019 the data implies $L_t/N_t$, changing from 8.0 to 5.6 employees per firm. The structural equation implies a series $\vartheta_t$, given $\{L_t/N_t,\phi, \alpha, \mu_t, \nu_t\}$, ranging from 5.4 to 14.6 and averaging $\vartheta=9.8$. Varying $\vartheta_t$ rather than taking the average does not affect our results meaningfully.

Our $\vartheta$ calibration satisfies the two theoretical assumptions we place on $\vartheta$. These ensure $A(\jmath)^{1/(\mu-\nu)}$ has a finite first moment and aggregate output is concave in aggregate capital. For our calibration the restrictions imply that we must set $\vartheta > 7$.

Over 2000--2019 $\mu_t-\nu_t$ lies between 0.139 and 0.274, averaging 1.8, which implies a scaled Pareto tail parameter $\vartheta^\ast \in (1.39,2.74)$ and an average $\vartheta^\ast=1.8$.

\subsubsection{Robustness Check}
As a robustness check or alternative calibration of the Pareto tail parameter $\vartheta$, we use data about the upper tail of the firm size distribution.

Employment is proportional to scaled productivity, $\ell\propto A^{1/(\mu-\nu)}$, so if productivity is Pareto distributed with shape parameter $\vartheta$, employment is Pareto with tail parameter $\vartheta^\ast\equiv\vartheta(\mu-\nu)$. For a Pareto variable $\ell$ with lower bound $\ell_0$, the conditional mean satisfies
$\mathbb{E}[\ell\mid\ell>\ell_0] =\vartheta^\ast/(\vartheta^\ast-1)\,\ell_0$. 

Using ONS business dynamism data \parencite{ONS2024_BusinessDynamism}, large firms (with more than 250 employees $\ell> \ell_0 = 250$), denoted set $\mathcal{M}$, accounted for $100\cdot \sum_{\imath \in \mathcal{M}} \ell_\imath / L = 52.4\%$ of employment in 2024 while representing only
$100\cdot \sum_{\imath \in \mathcal{M}} n_\imath / N =0.4\%$ of firms. This implies $\sum_{\imath \in \mathcal{M}} \ell_\imath / \sum_{\imath \in \mathcal{M}} n_\imath  \times (N/L)= 52.4/0.4 =131$. Given overall average firm employment of $L/N \approx 6$ workers, then average
employment in the $\ell>250$ bin is $\sum_{\imath \in \mathcal{M}} \ell_\imath / \sum_{\imath \in \mathcal{M}} n_\imath = 6 \times 131 = 786$. Setting $\ell_0=250$ therefore yields
$786=\vartheta^\ast/(\vartheta^\ast-1)\times250$, implying
$\vartheta^\ast\simeq1.46$. Given our estimate $\mu-\nu=0.18$, this corresponds to
$\vartheta\approx8.1$, slightly below our baseline calibration of $\vartheta=10$. 

Alternatively, this approach could be used to verify our $\vartheta=10$ calibration. Substituting our benchmark parameters into the conditional mean for large firms yields $\mathbb{E}[\ell\mid\ell>\ell_0]
=250 \times 0.18 \times 10 / (0.18 \times 10 - 1) = 250 \times 1.8/0.8 = 563$, lower than our target average of 786.

\setcounter{section}{0}
\renewcommand{\thesection}{\Roman{section}}
\clearpage
\pagenumbering{roman}
\setcounter{page}{1}

\numberwithin{equation}{section}
\numberwithin{figure}{section}
\numberwithin{table}{section}

\section*{
Supplementary Appendix
}

This appendix contains notes to supplement the main appendix.

\section{Scale Economies Background} \label{sec:SE_background}

In this section, we define some concepts that are occasionally subject to ambiguity. 

\textit{Internal vs. External Returns to Scale:} Our interest is in internal returns to scale, not external returns to scale that arise from aggregation. Internal returns to scale and scale economies arise within the firm from the production technology or fixed costs. External returns to scale are gains in aggregate output from changing aggregate inputs. They arise from grouping firms together.\footnote{On the demand-side, with a consumption aggregator, the analogous concept is love-of-variety. Other terms used are `thick markets' \parencite{CaballeroLyons1992_JME}, Ethier effects \parencite{Ethier1982_AER}, and agglomeration effects \parencite{Krugman1991_JPE}.} 

\textit{Scale Economies:} Scale economies describe the response of firm costs to output changes. They are measured by the inverse cost elasticity, which is the average cost to marginal cost ratio.\footnote{This definition of scale economies is common in industrial organization textbooks \parencite{Panzar1989_HoIO, ChurchWare2000_book, DavisGarces2009_book}, recent examples are \textcite{Syverson2019_JEP, ConlonMillerOtgonYao2023_AEApp}. It is sometimes recognised in macroeconomics, for example \textcite{RotembergWoodford1993_NBER, Basu2008_tNPDoE, BaqaeeFarhiSangani2023_Restud, LashkariBauerBoussard2024_AER}.}

\textit{Returns to scale:} Returns to scale are a property of the production technology. To be precise, they are captured by the degree of homogeneity of the production function. On the cost side, this parameter represents the slope of a firm's marginal cost curve.\footnote{Occasionally, researchers recognise this parameter as `span of control' since it is mathematically analogous to the span of control parameter in \textcite{Lucas1978_tBJoE}. In that context, it captures diminishing returns in managerial span of control. \textcite{Hopenhayn2014_ARE} analyses the equivalence with returns to scale.} For homothetic production functions, the scale elasticity of the cost function equals the returns to the scale of the production function.\footnote{\textcite[Ch. 8]{SilberbergSuen2000_book} present traditional proofs.} Fixed costs lead to non-homothetic production functions which break this relationship.

Imprecision over the terms scale economies and returns to scale extends beyond semantics. Erroneous conclusions and calibrations occur when the AC/MC ratio is estimated but is interpreted as the production function returns to scale.\footnote{\textcite{Basu2008_tNPDoE} discusses this in detail. Since homothetic production functions are common in macroeconomics, the term returns to scale is often used universally even in the presence of fixed costs.} 

\subsection{Graphical Intuition of Scale Economies}
To aid understanding throughout the paper, it is helpful to present the cost curve scenarios of the production functions we consider.
We define scale economies as the inverse cost elasticity, which is the ratio of average cost to marginal cost. With firm output $y$, we have:
\begin{equation*} \label{eq:rts_definition}
    S(y) \equiv \left(\frac{\partial \mathcal{C}}{\partial y} \frac{y}{\mathcal{C}} \right)^{-1} = \frac{AC(y)}{MC(y)}
\end{equation*}
where $AC \equiv \mathcal{C} / y$ and $MC \equiv \partial \mathcal{C} / \partial y$. There are economies of scale if $S(y) > 1$; constant scale economies if $S(y) = 1$; and diseconomies of scale if $S(y) < 1$.
Figure \ref{fig:firm_cost_curves_main} presents a firm with a U-shaped average cost curve due to increasing marginal costs and fixed cost.\footnote{In the appendix we present plots considering the three main cases that arise in our theory: a fixed cost with increasing, constant or decreasing marginal cost.} At the intersection of average and marginal cost, a firm has constant scale economies. To the left there are economies of scale. To the right there are diseconomies of scale. Therefore, the $S(y)$ curve shows that size and scale economies are negatively related at the firm level.\footnote{In Section \ref{sec:scale_econ_diags} we present a  graphical explanation of scale economies from the production side.}
\begin{figure}[H]
    \centering
    \begin{tikzpicture}[scale=1,thick]
            \def\a{0.2}
            \def\b{0.27}
            \def\w{1}
            \def\r{3.1}
            \def\intercept{60}
            \def\slope{-5}
            \def\FC{55.4}
            
            \begin{axis}[
            restrict y to domain=0:\intercept*0.5,
            samples = 100,
            xmin = 0, xmax = \intercept*0.5,
            ymin = 0, ymax = \intercept*1.3,
            xlabel = Output,
            ylabel = Costs $AC$ and $MC$,
            axis y line = left,
            axis x line = bottom,
            ticks=none
            ]
            \addplot[color=blue, mark=none, domain=2:\intercept*0.5, thick,name path=ATC] {\w*((\b*\w)/(\a*\r))^(-\b/(\a+\b))*\x^((1-\a-\b)/(\a+\b)) + ((\r*\FC)/\x)} node [pos=0.3,pin={-10:AC},inner sep=0pt] {};
            \addplot[color=red, mark=none, domain=2:\intercept*0.5, thick,name path=MC] {(\w/(\a+\b))*((\b*\w)/(\a*\r))^(-\b/(\a+\b))*\x^((1-\a-\b)/(\a+\b))}  node [pos=0.4,pin={-10:MC},inner sep=0pt] {};
            \addplot[color=cyan, mark=none, domain=2:\intercept*0.5, thick,name path=RTS] {(\w*((\b*\w)/(\a*\r))^(-\b/(\a+\b))*\x^((1-\a-\b)/(\a+\b)) + ((\r*\FC)/\x))/((\w/(\a+\b))*((\b*\w)/(\a*\r))^(-\b/(\a+\b))*\x^((1-\a-\b)/(\a+\b)))} node [pos=0.35,pin={45:S(y)},inner sep=0pt] {}; 
            \end{axis}
            \end{tikzpicture}
    \caption{Fixed Cost with Increasing MC, U-Shaped AC Curve}
    \label{fig:firm_cost_curves_main}
\end{figure}

\textit{Profits, Markups and Scale Economies:} Scale economies can be represented directly from the profit definition. This yields an expression based on market structure, namely markups and profits. Scale economies can also be written in terms of technical properties of the production function, namely fixed costs and the homogeneity parameter. This will depend on the production function and can be derived from the cost function or the production function.\footnote{In this paper we will show this for labour denominated fixed costs beginning with the production function. \textcite{Savagar2021_JEDC} shows it for output-denominated fixed costs beginning with the cost function.} Consider the definition of profits as revenue minus costs
\begin{equation*}
    \text{Profit} = \text{Price} \times \text{Output} - \text{Cost} = \text{Revenue} - \text{Cost}. 
\end{equation*}
Divide by revenue, define AC=Cost/Output, and multiply by MC/MC, yields:
\begin{equation*}
    \frac{\text{AC}}{\text{MC}} = \frac{\text{Price}}{\text{Marginal Cost}} \left(1 - \frac{\text{Profit}}{\text{Revenue}} \right).
\end{equation*}
This shows that a firm's scale economies are its markup multiplied by its profit share remainder (\textit{i.e.} total cost share).\footnote{The total cost share is the sum of the variable cost share and the fixed cost share.} A firm that makes zero-profits has scale economies equal to its markup.\footnote{This result was used in earlier empirical work on returns to scale, when profits in the US economy were close to zero \parencite{BasuFernald1997_JPE}.} And, a firm with positive profits will have lower scale economies than the zero-profit firm. Higher scale economies imply higher markups or lower profit shares. Since we develop a framework with constant markups, differences in scale economies are analagous to differences in profits shares. Large, high-productivity, firms have large profit shares and low scale economies, whilst small, low-productivity, firms have low profit shares and high scale economies. 

Figure \ref{fig:large_v_small_scale_economies} illustrates scale economies from the production side. It conveys the idea that small firms have high scale economies, whilst large firms have low scale economies. The figure represents an economy where firm output is produced directly by production labour. In order to produce there is some overhead labour that is the same for both firms. Total labour is the sum of production labour and overhead labour. The figure shows that a 10\% rise in total labour at a firm raises production labour by 100\% for the small firm, but only 13\% for the large firm. Therefore, a proportional change in inputs has a proportionally larger effect on output for the small firm.
\pgfplotstableread{
Label     Total Production Overhead
Small    1  1 9
Large    4  31 9
}\testdata
\begin{figure}[H]
\centering
\begin{tikzpicture}
    \begin{axis}[
    ybar stacked,
    bar width=20pt,
    ymin=0,
    xtick=data,
    legend style={at={(axis cs:0.2,20)},anchor=south west},
    xticklabels from table={\testdata}{Label},
    yticklabel style={name=T\ticknum}
    ]
    \draw[green!80, dashed] (axis cs:-1, 9)--(axis cs:2,9) node [midway, below] {9};
    \draw[blue!60, dashed] (axis cs:-1,10)--(axis cs:2,10) node [midway, above] {10};
    \draw[blue!60, dashed] (axis cs:-1,40)--(axis cs:2,40) node [midway, above] {40};
    \addplot [fill=green!80] table [y=Overhead, meta=Label,x expr=\coordindex] {\testdata};
    \addplot [fill=blue!60] table [y=Production, meta=Label,x expr=\coordindex] {\testdata};
    \addplot [fill=red!60]
    table [y=Total, meta=Label,x expr=\coordindex] {\testdata};
    \legend{Overhead Labour,Production Labour,10\% Rise Total}
    \end{axis}
\end{tikzpicture}
    \caption{Scale Economies for Large and Small Firm}
    \label{fig:large_v_small_scale_economies}
\end{figure}

\subsection{Graphical Illustration of Scale Economies (Cost based)}\label{sec:scale_econ_diags}
It is helpful to consider the three types of cost curve scenarios faced by firms in our model.

Figures \ref{fig:firm_cost_curves1}, \ref{fig:firm_cost_curves2} and \ref{fig:firm_cost_curves3} show a firm's cost curves for the case where there is a fixed cost and increasing, constant or decreasing marginal costs. The diagrams show average total cost (ATC), average variable cost (AVC), average fixed cost (AFC) and marginal cost (MC) as firm output varies. Specifically, total cost is the sum variable cost and a fixed cost:
$TC = VC + FC$, and averages are the components when divided by output $y$.
The demand curve ($p(y)$) and marginal revenue (MR) curve ($\frac{d \; p(y) y}{d y}$) are not shown. We can imagine them as horizontal in the perfectly competitive case and downward sloping with imperfect competition, for example, due to product differentiation. The first case (Figure \ref{fig:firm_cost_curves1}) allows for a perfectly competitive equilibrium when the demand curve is horizontal and firms produce at minimum average cost. The second and third cases (Figure \ref{fig:firm_cost_curves2} and \ref{fig:firm_cost_curves3}) require imperfect competition. The demand curve must be downward sloping for $MR=MC$ to occur.

Figure \ref{fig:firm_cost_curves1} illustrates the cost curves of a firm with a fixed cost and increasing marginal cost curve. The firm's marginal cost intersects the average total cost at its minimum. This minimum point is the firm's \textit{minimum efficient scale} (MES) which would arise under perfect competition and at this minimum the firm has constant scale. To the left-hand side of the MES the firm has economies of scale and to the right-hand side the firm has diseconomies of scale.
\begin{figure}[H]
    \centering
    \begin{tikzpicture}[scale=1,thick]
            \def\a{0.2}
            \def\b{0.27}
            \def\w{1}
            \def\r{3.1}
            \def\intercept{60}
            \def\slope{-5}
            \def\FC{55.4}
            
            \definecolor{amethyst}{rgb}{0.6, 0.4, 0.8}
            \definecolor{blush}{rgb}{0.87, 0.36, 0.51}
            
            \begin{axis}[
            ticks=none,
            restrict y to domain=0:\intercept*0.5,
            samples = 100,
            xmin = 0, xmax = \intercept*0.5,
            ymin = 0, ymax = \intercept*1.3,
            xlabel = Output $(y)$,
            ylabel = Costs $AC(y)$ and $MC(y)$,
            axis y line = left,
            axis x line = bottom,
            y axis line style = {-},
            x axis line style = {-}
            ]

            \addplot[color=blue, mark=none, domain=2:\intercept*0.5, thick,name path=ATC] {\w*((\b*\w)/(\a*\r))^(-\b/(\a+\b))*\x^((1-\a-\b)/(\a+\b)) + ((\r*\FC)/\x)} node [pos=0.3,pin={-10:ATC},inner sep=0pt] {};
            \addplot[color=red, mark=none, domain=2:\intercept*0.5, thick,name path=MC] {(\w/(\a+\b))*((\b*\w)/(\a*\r))^(-\b/(\a+\b))*\x^((1-\a-\b)/(\a+\b))}  node [pos=0.4,pin={-10:MC},inner sep=0pt] {};
            \addplot[color=black, mark=none, domain=2:\intercept*0.5, thick,name path=AVC] {\w*((\b*\w)/(\a*\r))^(-\b/(\a+\b))*\x^((1-\a-\b)/(\a+\b))} node [pos=0.7,pin={-10:AVC},inner sep=0pt] {}; 
            \addplot[color=brown, mark=none, domain=2:\intercept*0.5, thick,name path=AFC] {(\r*\FC)/\x} node [pos=0.9,pin={45:AFC},inner sep=0pt] {}; 
            \end{axis}
            \end{tikzpicture}
    \caption{Fixed Cost with Increasing MC, U-Shaped AC Curve}
    \label{fig:firm_cost_curves1}
\end{figure}
Figure \ref{fig:firm_cost_curves2} has a constant marginal cost curve and a fixed cost, so there are globally decreasing returns and ATC=MC in the limit. In this case there must be a downward sloping demand curve for an equilibrium where $MR=MC$ to exist. Any degree of slope in the demand curve is sufficient to give an equilibrium, unlike in the next example example which requires a sufficiently steep demand curve (or a sufficiently shallow decreasing marginal cost).
\begin{figure}[H]
    \centering
    \begin{tikzpicture}[scale=1,thick]
            \def\a{0.7}
            \def\b{0.3}
            \def\w{1}
            \def\r{3.1}
            \def\intercept{10}
            \def\slope{-5}
            \def\FC{5}
            
            \definecolor{amethyst}{rgb}{0.6, 0.4, 0.8}
            \definecolor{blush}{rgb}{0.87, 0.36, 0.51}
            
            \begin{axis}[             ticks=none,
            restrict y to domain=0:\intercept*0.5,
            samples = 100,
            xmin = 0, xmax = \intercept*0.5,
            ymin = 0, ymax = \intercept*1.3,
            xlabel = Output $(y)$,
            ylabel = Costs $AC(y)$ and $MC(y)$,
            axis y line = left,
            axis x line = bottom,
            y axis line style = {-},
            x axis line style = {-}
            ]
            \addplot[color=blue, mark=none, domain=0.5:\intercept*0.5, thick,name path=ATC] {\w*((\b*\w)/(\a*\r))^(-\b/(\a+\b))*\x^((1-\a-\b)/(\a+\b)) + ((\r*\FC)/\x)} node [pos=0.9,pin={45:ATC},inner sep=0pt] {};
            \addplot[color=red, mark=none, domain=0.5:\intercept*0.5, thick,name path=MC] {(\w/(\a+\b))*((\b*\w)/(\a*\r))^(-\b/(\a+\b))*\x^((1-\a-\b)/(\a+\b))}  node [pos=0.4,pin={-10:MC},inner sep=0pt] {};
            \addplot[color=black, mark=none, domain=0.5:\intercept*0.5, thick,name path=AVC] {\w*((\b*\w)/(\a*\r))^(-\b/(\a+\b))*\x^((1-\a-\b)/(\a+\b))} node [pos=0.7,pin={-10:AVC},inner sep=0pt] {}; 
            \addplot[color=brown, mark=none, domain=0.5:\intercept*0.5, thick,name path=AFC] {(\r*\FC)/\x} node [pos=0.9,pin={225:AFC},inner sep=0pt] {}; 
            \end{axis}
            \end{tikzpicture}
    \caption{Fixed Cost with Constant MC, Globally Decreasing Returns}
    \label{fig:firm_cost_curves2}
\end{figure}
Figure \ref{fig:firm_cost_curves3} has a decreasing marginal cost and a fixed cost so there are global diseconomies of scale. In this case there must be a downward sloping demand curve for an equilibrium where $MR=MC$ to exist. The demand curve must be steeper than the downward-sloping marginal cost curve to ensure this occurs.
\begin{figure}[H]
    \centering
    \begin{tikzpicture}[scale=1,thick]
            \def\a{1.2}
            \def\b{0.8}
            \def\w{1}
            \def\r{3.1}
            \def\intercept{9}
            \def\slope{-5}
            \def\FC{1}
            
            \definecolor{amethyst}{rgb}{0.6, 0.4, 0.8}
            \definecolor{blush}{rgb}{0.87, 0.36, 0.51}
                        \begin{axis}[             ticks=none,
            restrict y to domain=0:\intercept*0.5,
            samples = 100,
            xmin = 0, xmax = \intercept*0.5,
            ymin = 0, ymax = \intercept*1.1,
            xlabel = Output $(y)$,
            ylabel = Costs $AC(y)$ and $MC(y)$,
            axis y line = left,
            axis x line = bottom,
            y axis line style = {-},
            x axis line style = {-}
            ]
            \addplot[color=blue, mark=none, domain=0.5:\intercept*0.5, thick,name path=ATC] {\w*((\b*\w)/(\a*\r))^(-\b/(\a+\b))*\x^((1-\a-\b)/(\a+\b)) + ((\r*\FC)/\x)} node [pos=0.9,pin={80:ATC},inner sep=0pt] {};
            \addplot[color=red, mark=none, domain=0.5:\intercept*0.5, thick,name path=MC] {(\w/(\a+\b))*((\b*\w)/(\a*\r))^(-\b/(\a+\b))*\x^((1-\a-\b)/(\a+\b))}  node [pos=0.2,pin={160:MC},inner sep=0pt] {};
            \addplot[color=black, mark=none, domain=0.5:\intercept*0.5, thick,name path=AVC] {\w*((\b*\w)/(\a*\r))^(-\b/(\a+\b))*\x^((1-\a-\b)/(\a+\b))} node [pos=0.2,pin={150:AVC},inner sep=0pt] {}; 
            \addplot[color=brown, mark=none, domain=0.5:\intercept*0.5, thick,name path=AFC] {(\r*\FC)/\x} node [pos=0.3,pin={45:AFC},inner sep=0pt] {}; 
            \end{axis}
            \end{tikzpicture}
    \caption{Fixed Cost with Decreasing MC, Globally Decreasing Returns}
    \label{fig:firm_cost_curves3}
\end{figure}

\subsection{Scale Economies (model-based representation)}
The firm-level production function is $y_t(\jmath) = A_t(\jmath) k_t(\jmath)^{\alpha \nu} \ell_t(\jmath)^{(1-\alpha)\nu}$, where production labour is total labour minus overhead labour: $\ell_t(\jmath) = \ell_t^{\mathrm{tot}} - \phi$. The parameters $\nu$ and $\phi$ are both sources of scale economies in the model. That is, they both affect the response of costs to output changes, equivalently, the response of output to \textit{all} input changes. Scale economies are measured as the ratio of average cost to marginal cost (the inverse cost elasticity to output).

Below, we derive firm-level scale economies from the production side by summing output elasticities. The same result can be shown from the firm-level cost function.\footnote{\textcite{Savagar2021_JEDC} shows this for a model with output denominated fixed costs.}

The response of firm output to a change in each variable input is constant. Consequently, returns to scale in variable inputs is constant:
\begin{align*}
    \frac{\partial \ln y_t(\jmath)}{\partial \ln k_t(\jmath)} = \nu \alpha
    , \quad 
    \frac{\partial \ln y_t(\jmath)}{\partial \ln \ell_t(\jmath)} = \nu (1-\alpha), \quad \frac{\partial \ln y_t(\jmath)}{\partial \ln k_t(\jmath)} + \frac{\partial \ln y_t(\jmath)}{\partial \ln \ell_t (\jmath)} = \nu.
\end{align*}
The effect of a change in total labour input is decreasing in firm size:\footnote{For the second equality, we use the zero-profit condition $\left(1 - \frac{\nu}{\mu} \right)p_t(\jmath)y_t(\jmath) = w_t \phi \left( \frac{A(\jmath)}{\underbar{A}_t}\right)^{\frac{1}{\mu - \nu}}$ combined with labour demand $\frac{w_t}{p_t(\jmath)y_t(\jmath)} = \frac{\nu(1-\alpha)}{\mu} \frac{1}{\ell(\jmath)}$ to yield $\nu(1-\alpha) \frac{\phi}{\ell_t(\jmath)} = (\mu - \nu)\left( \frac{\underbar{A}_t}{A(\jmath)}\right)^{\frac{1}{\mu - \nu}} $.}
\begin{equation*}
    \frac{\partial \ln y_t(\jmath)}{\partial \ln \ell_t^{\mathrm{tot}}(\jmath)} = \nu (1-\alpha) \left(1 + \frac{\phi}{\ell_t(\jmath)} \right) = \nu(1-\alpha) + (\mu - \nu) \left( \frac{\underbar{A}_t}{A(\jmath)} \right)^{\frac{1}{\mu - \nu}} \quad \in \left(\nu(1-\alpha), \mu - \alpha \nu \right).
\end{equation*}
Therefore, scale economies at the firm are decreasing in firm size:
\begin{equation} \label{eq:rts_labden}
    S_t(\jmath) \equiv  \frac{\partial \ln y_t(\jmath)}{\partial \ln k_t(\jmath)} + \frac{\partial \ln y_t(\jmath)}{\partial \ln \ell_t^{tot}(\jmath)} = \nu\left( 1 + (1-\alpha) \frac{\phi}{\ell_t(\jmath)} \right) = \nu + (\mu - \nu)\left(\frac{\underbar{A}_t}{A(\jmath)}\right)^{\frac{1}{\mu - \nu}} \quad \in \left(\nu, \mu \right).
\end{equation}
A firm's scale economies decrease as production labour rises relative to the labour overhead, or as firm productivity rises relative to the productivity cut-off. Figure \ref{fig:rts_schedule_baseline} plots \eqref{eq:rts_labden} for a given $\underbar{A}$. More productive firms have lower scale economies. The cut-off firm has the highest level of scale equals to the markup, and scale converges on returns to scale in variable inputs $\nu$ for high-productivity firms.
\begin{figure}[H]
    \centering
    \singlespacing
    \caption{Firm-level Scale Economies in Steady-State}
    \label{fig:rts_schedule_baseline}
    \pgfmathsetmacro{\muval}{1.1}
\pgfmathsetmacro{\nuval}{0.9}
\pgfmathsetmacro{\vartheta}{1.3}
\pgfmathsetmacro{\Jval}{0.081178}
\pgfmathsetmacro{\domainmax}{6}
\pgfmathsetmacro{\Abar}{pow((\vartheta/(\vartheta-1))*pow(1/(1-\Jval),1/\vartheta), (1-\nuval))}

\begin{tikzpicture}
\begin{axis}[
  axis lines=left,
  width=0.9\textwidth,
  height=\axisdefaultheight,
  xlabel={$A(\jmath)$},
  ylabel={$S(\jmath)$},
  xmin=0,
  xmax=1.1*\domainmax,
  ymin=0.7,
  ymax=1.3,
  extra y ticks={\muval,\nuval},
  extra y tick labels={$\mu$,$\nu$},
  extra x ticks={\Abar},
  extra x tick labels={$\underbar{A}$},
  yticklabels=\empty,
  xticklabels=\empty,
  ytick style={draw=none},
  xtick style={draw=none},
]

\addplot[
  black,
  domain=0.5:\domainmax,
  samples=200,
  name path=Scurve
] { \nuval + (\muval-\nuval)*pow(\Abar/x, 1/(\muval-\nuval)) }
  node[anchor=south, pos=0.92, inner xsep=0pt] {$S(\jmath)$};

\addplot[
  black,
  dashed,
  domain=0.5:\Abar,
  samples=200
] { \nuval + (\muval-\nuval)*pow(\Abar/x, 1/(\muval-\nuval)) };

\addplot[gray!75, dashed, domain=0:\domainmax] {\nuval};
\addplot[gray!75, dashed, domain=0:\domainmax] {\muval};

\addplot[
  draw=none,
  name path=Abarline
]
coordinates {
  (\Abar,\pgfkeysvalueof{/pgfplots/ymin})
  (\Abar,\pgfkeysvalueof{/pgfplots/ymax})
};

\path[name path=yaxis]
  (\pgfkeysvalueof{/pgfplots/xmin},\pgfkeysvalueof{/pgfplots/ymin})
  -- (\pgfkeysvalueof{/pgfplots/xmin},\pgfkeysvalueof{/pgfplots/ymax});

\addplot[
  gray!50,
  fill opacity=0.5
]
fill between[of=yaxis and Abarline];

\end{axis}
\end{tikzpicture}
    \caption*{\footnotesize Plot shows equation \eqref{eq:rts_labden} scale of a firm given its productivity draw. In the shaded region firms are inactive and the dashed line shows their hypothetical scale economies if they were to produce. 
    The horizontal lines show the bounds on scale economies of active firms $S(\jmath) \in (\nu,\mu)$. We have assumed $A(\jmath)$ is Pareto distribution and we have set $\underbar{A}$ arbitrarily.
    }
\end{figure}

\section{Pareto Distributed Productivity}
We obtain a measure of productivity $A(\jmath)$ from a random draw on the unit interval $\jmath \in [0,1]$ using inverse transform sampling. 
The Pareto CDF is given by \[F(A; \vartheta) = 1-\left(\frac{h}{A}\right)^{\vartheta}; \quad A \geq h >0 \quad \text{and} \quad \vartheta>0.\]  If $\mathcal{J} \sim Uniform(0,1]$, then for $\jmath \in \mathcal{J}$, we have \[1-\left(\frac{h}{A}\right)^{\vartheta} = \jmath\] Therefore, the \textit{quantile function} is \[A(\jmath) = h(1-\jmath)^{-\frac{1}{\vartheta}}.\] Typically we set the scale parameter, which is the minimum possible value of $A$, to $h=1$. Calibrations of the shape parameter (tail index) vary, for example $\vartheta = 1.15$ in \textcite{BarseghyanDiCecio2011_JET} and $\vartheta = 1.06$ in \textcite{Luttmer2007_QJE} and $\vartheta=6.10$ in \textcite{AsturiasHurKehoeRuhl2023_AEJM}. These estimates are set to match the firm size distribution in terms of employment since in these models $A(\jmath)$ is proportional to employment, though, as below, scaling can affect this.
\begin{figure}[H]
    \centering
    \begin{tikzpicture}
    \begin{axis}[
            xlabel = $\jmath$,
            ylabel = $A(\jmath)$,
            ymin = 0,
            xmin= 0,
            xmax = 1,
            ]
    \addplot[black, thick, domain=0:0.97, samples=100]{(1-x)^(-1/6.10)};
    \addlegendentry{$A(\jmath) = (1-\jmath)^{-\frac{1}{6.10}}$}
    \addplot[blue, thick, domain=0:0.97, samples=100]{(1-x)^(-1/1.15)};
    \addlegendentry{$A(\jmath) = (1-\jmath)^{-\frac{1}{1.15}}$}
    \addplot[red, thick, domain=0:0.97, samples=100]{(1-x)^(-1/1.06)};
    \addlegendentry{$A(\jmath) = (1-\jmath)^{-\frac{1}{1.06}}$}
    \end{axis}
    \end{tikzpicture}    
    \caption{Productivity with Pareto Distribution, $h=1, \vartheta=\{1.06, 1.15\}$. Domain $\jmath \in (0:0.97)$}
    \label{fig:Tech_Pareto_distribution}
\end{figure}
Figure \ref{fig:Scaled_Tech_Pareto_distribution_changing_nu} plots scaled technology $A(\jmath)^{\frac{1}{\mu - \nu}}$ for different calibrations of $\nu=\{0.95,1.00,1.05\}$ given fixed values of $\mu=1.1$ and $\vartheta=50$. This changes the effective tail index: $(\mu-\nu)\vartheta$. The distribution of scaled technology is proportional to the distribution of labour, capital and revenue. We require that the effective tail index exceeds one, $(\mu - \nu) \vartheta > 1$, so that the expected value of scaled technology is finite, and consequently the expected value of labour per firm, capital per firm and revenue per firm is not infinite.

A higher $\nu$ or lower $\mu$ leads to a lower effective tail index and greater scaled technology for any given $\jmath$ draw. A lower effective tail index causes a lower density of firms to have low-productivity draws and a greater density of firms to have high-productivity draws. Since employment, capital, and revenue are proportional to this, it also means the distribution of firms is denser towards large firms in terms of labour, capital and employment.
\begin{figure}[H]
    \centering
    \begin{tikzpicture}
    \begin{axis}[
            xlabel = $\jmath$,
            ylabel = $A(\jmath)^{\frac{1}{\mu-\nu}}$,
            xmin= 0,
            xmax = 1,
            ]
    \addplot[black, thick, domain=0:0.99, samples=100]{(1-x)^(-1/((1.1-0.95)*50))};
    \addlegendentry{$A(\jmath)^{\frac{1}{\mu-\nu}} = (1-\jmath)^{-\frac{1}{(\mu-0.95)\vartheta}}$}
    \addplot[blue, thick, domain=0:0.99, samples=100]{(1-x)^(-1/((1.1-1.00)*50))};
    \addlegendentry{$A(\jmath)^{\frac{1}{\mu-\nu}} = (1-\jmath)^{-\frac{1}{(\mu-1)\vartheta}}$}
    \addplot[red, thick, domain=0:0.99, samples=100]{(1-x)^(-1/((1.1-1.05)*50))};
    \addlegendentry{$A(\jmath)^{\frac{1}{\mu-\nu}} = (1-\jmath)^{-\frac{1}{(\mu-1.05)\vartheta}}$}
    \end{axis}
    \end{tikzpicture}    
    \caption{Scaled Technology with Pareto Distribution, $h=1, \vartheta=50$ and $\mu=1.1, \; \nu = \{0.95,1.00,1.05\}$. Domain $\jmath \in (0:0.95)$.}
    \label{fig:Scaled_Tech_Pareto_distribution_changing_nu}
\end{figure}

\section{Additional Model Derivations}

\subsection{Profit Maximization Problem}
\label{app:profit_max_conditions}

\subsubsection*{First-Order Conditions}
We drop time subscripts $t$ and firm-specific notation $\jmath$. Fixed parameters are $\{\nu, \mu, \alpha, \phi\}$ and endogenous variables are $\{N,Y,A,k,\ell,r,w\}$. The revenue function is 
\begin{align*}
    py = N^{\frac{1 - \mu}{\mu}} Y^{\frac{\mu - 1}{\mu}} y^{\frac{1}{\mu}} = N^{\frac{1 - \mu}{\mu}} Y^{\frac{\mu - 1}{\mu}}A^{\frac{1}{\mu}}k^{\frac{\alpha \nu}{\mu}} \ell^{\frac{(1 - \alpha)\nu}{\mu}}.
\end{align*}
The variables $\{N,Y,A, w, r\}$ are taken as given by the firm. The firm maximizes revenue less costs:
\begin{align*}
  \max_{k, \ell} \quad  p(k,\ell)y(k,\ell) - r k - w(\ell+\phi) .
\end{align*}
The first-order conditions of the maximization problem state that the marginal revenue product of labour (MRPL) and marginal revenue product of capital (MRPK) -- \textit{i.e.} the revenue derivatives with respect to labour and capital -- equal to wage and rental rate at optimal choices:
\begin{align*}
    MRPL = \frac{\nu (1-\alpha)}{\mu} \frac{p(k^*, \ell^*) y(k^*, \ell^*)}{\ell^*} &=  w
    \\
     MRPK = \frac{\nu\alpha}{\mu} \frac{p(k^*, \ell^*) y(k^*, \ell^*)}{k^*}   &= r.    
\end{align*}
Since $0 < \alpha <1$, $\mu \geq 1$, $\nu>0$ the marginal revenue products are positive. Asterisk notation denotes the profit-maximizing levels of capital and labour.

\subsubsection*{Second-Order Conditions}

The second-order conditions for maximization require that, at the optimal point $\{k^*, \ell^*\}$, the objective function is decreasing in capital and labour and the determinant of the Hessian of the objective function is positive. This implies that $MRPL_\ell < 0$ and $MRPK_k < 0$ where subscripts denote derivatives. And, $MRPL_\ell MRPK_k - MRPL_k^2  >0$. First note:
\[    MRPL_k = MRPK_\ell = \frac{\nu \alpha}{\mu} \frac{MRPL}{k^*} = \frac{\nu(1-\alpha)}{\mu} \frac{MRPK}{\ell^*}.
\]
Therefore the following conditions must be satisfied:
\begin{align*}
    MRPL_\ell &= \left( \frac{\nu (1-\alpha)}{\mu} - 1\right)\frac{MRPL}{\ell^*} < 0 
    \\
    MRPK_k &= \left( \frac{\nu \alpha}{\mu} - 1\right)\frac{MRPK}{k^*} < 0 
    \\
    MRPL_\ell MRPK_k - MRPL_k^2 & =  \frac{MRPL \times MRPK}{k^* \ell^*} \left( 1 - \frac{\nu}{\mu} \right) > 0
\end{align*}
These conditions hold if $\nu<\mu$. 

\subsection{Reduced-form Aggregate Output}

Our aggregate output expression generalises several examples in the literature and is similar to \textcite{BarseghyanDiCecio2016_EER}. Our result simplifies to the result in \textcite{KehrigGao2021_wp} when there is perfect competition $\mu=1$, or with imperfect competition $\mu = \frac{\theta}{\theta -1}>1$ but constant marginal cost $\nu=1$ we have productivity as defined in \textcite{Melitz2003_ecta, GhironiMelitz2005_QJE}.

Aggregate output can be written as a Cobb--Douglas function of aggregate capital and labour, scaled by a power-mean measure of firm-level productivity. Starting from the final-goods aggregator,
\begin{align}
Y_t
&= N_t\!\left[\frac{1}{N_t}\int_0^{N_t} y_t(\imath)^{\frac{1}{\mu}}\,d\imath\right]^{\mu}
= N_t\!\left[\frac{E_t}{N_t}\int_{J_t}^{1} y_t(\jmath)^{\frac{1}{\mu}}\,d\jmath\right]^{\mu}
= N_t\!\left[\frac{1}{1-J_t}\int_{J_t}^{1} y_t(\jmath)^{\frac{1}{\mu}}\,d\jmath\right]^{\mu}.
\end{align}

We express firm-level output relative to the threshold firm. Since $y_t(J_t)$ is constant across $\jmath$, this yields
\begin{align}
Y_t
= N_t\,y_t(J_t)
\left[\frac{1}{1-J_t}\int_{J_t}^{1}
\left(\frac{y_t(\jmath)}{y_t(J_t)}\right)^{\frac{1}{\mu}} d\jmath
\right]^{\mu}.
\end{align}

Using firms’ optimal pricing and input choices, relative output satisfies
\begin{equation}
\left(\frac{y_t(\jmath)}{y_t(J_t)}\right)^{\frac{1}{\mu}}
=
\frac{p_t(\jmath)y_t(\jmath)}{p_t(J_t)y_t(J_t)}
=
\left(\frac{A(\jmath)}{\underline A_t}\right)^{\frac{1}{\mu-\nu}}.
\end{equation}
Hence,
\begin{align}
Y_t
&= N_t\,y_t(J_t)
\left[
\frac{1}{1-J_t}\int_{J_t}^{1}
\left(\frac{A(\jmath)}{\underline A_t}\right)^{\frac{1}{\mu-\nu}}
d\jmath
\right]^{\mu}
= N_t\,y_t(J_t)
\left(\frac{\hat A_t}{\underline A_t}\right)^{\frac{\mu}{\mu-\nu}}.
\end{align}

Aggregate output therefore depends on the number of active firms, the scale of the marginal firm, and the productivity gap between the average incumbent and the cutoff firm.\footnote{\textcite{KehrigGao2021_wp} derive an analogous result in partial equilibrium under perfect competition ($\mu=1$).}

Substituting the threshold firm’s production function,
\[
y_t(J_t)=\underline A_t\big[k_t(J_t)^{\alpha}\ell_t(J_t)^{1-\alpha}\big]^{\nu},
\]
gives
\begin{equation}
Y_t
= N_t\,\hat A_t^{\frac{\mu}{\mu-\nu}}
\underline A_t^{-\frac{\nu}{\mu-\nu}}
\big[k_t(J_t)^{\alpha}\ell_t(J_t)^{1-\alpha}\big]^{\nu}.
\end{equation}

We next express threshold-firm inputs in terms of aggregates. Aggregate capital is
\begin{align}
K_t
&= E_t\int_{J_t}^{1}k_t(\jmath)\,d\jmath
= \frac{N_t}{1-J_t}\int_{J_t}^{1}k_t(\jmath)\,d\jmath
= \frac{N_t k_t(J_t)}{1-J_t}
\int_{J_t}^{1}\left(\frac{A(\jmath)}{\underline A_t}\right)^{\frac{1}{\mu-\nu}}d\jmath
\\
&= N_t\,k_t(J_t)
\left(\frac{\hat A_t}{\underline A_t}\right)^{\frac{1}{\mu-\nu}}.
\end{align}
Similarly, aggregate labour satisfies
\begin{align}
L_t
&= E_t\int_{J_t}^{1}\!\big[\ell_t(\jmath)+\phi\big]\,d\jmath
= N_t\,\ell_t(J_t)
\left(\frac{\hat A_t}{\underline A_t}\right)^{\frac{1}{\mu-\nu}}
+ N_t\phi.
\end{align}
Defining the fraction of labour used in production,
\[
u_t \equiv 1-\frac{N_t\phi}{L_t},
\]
we obtain
\begin{align}
k_t(J_t)
&=\left(\frac{\underline A_t}{\hat A_t}\right)^{\frac{1}{\mu-\nu}}\frac{K_t}{N_t},
\\
\ell_t(J_t)
&=\left(\frac{\underline A_t}{\hat A_t}\right)^{\frac{1}{\mu-\nu}}\frac{u_t L_t}{N_t}.
\end{align}

Substituting back yields the reduced-form aggregate production function:
\begin{equation}
\boxed{
Y_t
= N_t^{1-\nu}\,\hat A_t
\left[K_t^{\alpha}(u_t L_t)^{1-\alpha}\right]^{\nu}.
}
\end{equation}

\section{General Comparative Statics without the \texorpdfstring{$u$}{u} Channel}
\label{app:comparative_statics_no_u}

This appendix derives general (out-of-steady-state and distribution-free) comparative statics when the labour allocation channel $u_t$ is substituted out. In this case, factor shares depend only on the transformed productivity gap $\Gamma_t^{1/(\mu-\nu)}$ and structural parameters.

\subsection{Factor shares}

Substituting out $u_t$ yields
\begin{align}
s_{\phi,t}
&= \left(1-\frac{\nu}{\mu}\right)\Gamma_t^{-\frac{1}{\mu-\nu}}, \label{eq:sphi_raw} \\
s_{L,t}
&= (1-\alpha)\frac{\nu}{\mu}
+ \left(1-\frac{\nu}{\mu}\right)\Gamma_t^{-\frac{1}{\mu-\nu}}, \label{eq:sL_raw} \\
s_{\pi,t}
&= \left(1-\frac{\nu}{\mu}\right)
\left(1-\Gamma_t^{-\frac{1}{\mu-\nu}}\right). \label{eq:spi_raw}
\end{align}
Define the transformed productivity gap
\begin{equation}
z_t \;\equiv\; \Gamma_t^{1/(\mu-\nu)} > 1.
\end{equation}
The factor shares can then be written compactly as
\begin{align}
s_{\phi,t}
&= \Big(1-\frac{\nu}{\mu}\Big) z_t^{-1}, \label{eq:sphi} \\
s_{L,t}
&= (1-\alpha)\frac{\nu}{\mu}
+ \Big(1-\frac{\nu}{\mu}\Big) z_t^{-1}, \label{eq:sL} \\
s_{\pi,t}
&= \Big(1-\frac{\nu}{\mu}\Big)\Big(1-z_t^{-1}\Big). \label{eq:spi}
\end{align}

\subsection{Overall comparative statics}

Let $x$ denote any underlying parameter (e.g.\ $x\in\{\mu,\nu,\phi,\ldots\}$). The \emph{total} derivative of any share $s$ with respect to $x$ decomposes into a direct effect (holding $z_t$ fixed) and an indirect effect operating through $z_t$:
\begin{equation}
\frac{d s}{d x}
=
\left.\frac{\partial s}{\partial x}\right|_{z_t}
+
\frac{\partial s}{\partial z_t}\frac{d z_t}{d x}.
\label{eq:total_derivative}
\end{equation}

\paragraph{Overhead share.}
From \eqref{eq:sphi},
\begin{equation}
\boxed{
\frac{d s_{\phi,t}}{d x}
=
\underbrace{
\left.\frac{\partial}{\partial x}
\Big(1-\frac{\nu}{\mu}\Big)\right|_{z_t} z_t^{-1}
}_{\text{direct}}
\;-\;
\underbrace{
\Big(1-\frac{\nu}{\mu}\Big) z_t^{-2}
\frac{d z_t}{d x}
}_{\text{via } z_t}
}
\end{equation}
with
\begin{equation}
\left.\frac{\partial}{\partial \mu}
\Big(1-\frac{\nu}{\mu}\Big)\right|_{z_t}
= \frac{\nu}{\mu^2} > 0,
\qquad
\left.\frac{\partial}{\partial \nu}
\Big(1-\frac{\nu}{\mu}\Big)\right|_{z_t}
= -\frac{1}{\mu} < 0.
\end{equation}

\paragraph{Total labour share.}
From \eqref{eq:sL},
\begin{equation}
\boxed{
\frac{d s_{L,t}}{d x}
=
\underbrace{
(1-\alpha)
\left.\frac{\partial}{\partial x}
\Big(\frac{\nu}{\mu}\Big)\right|_{z_t}
+
\left.\frac{\partial}{\partial x}
\Big(1-\frac{\nu}{\mu}\Big)\right|_{z_t} z_t^{-1}
}_{\text{direct}}
\;-\;
\underbrace{
\Big(1-\frac{\nu}{\mu}\Big) z_t^{-2}
\frac{d z_t}{d x}
}_{\text{via } z_t}
}
\end{equation}
where
\begin{equation}
\left.\frac{\partial}{\partial \mu}
\Big(\frac{\nu}{\mu}\Big)\right|_{z_t}
= -\frac{\nu}{\mu^2} < 0,
\qquad
\left.\frac{\partial}{\partial \nu}
\Big(\frac{\nu}{\mu}\Big)\right|_{z_t}
= \frac{1}{\mu} > 0.
\end{equation}

\paragraph{Profit share.}
From \eqref{eq:spi},
\begin{equation}
\boxed{
\frac{d s_{\pi,t}}{d x}
=
\underbrace{
\left.\frac{\partial}{\partial x}
\Big(1-\frac{\nu}{\mu}\Big)\right|_{z_t}
\Big(1-z_t^{-1}\Big)
}_{\text{direct}}
\;+\;
\underbrace{
\Big(1-\frac{\nu}{\mu}\Big) z_t^{-2}
\frac{d z_t}{d x}
}_{\text{via } z_t}
}
\end{equation}

\subsection{Sign implications conditional on \texorpdfstring{$dz_t/dx$}{dz/dx}}

Since $\mu>\nu$ implies $1-\nu/\mu>0$ and $z_t>1$ implies $z_t^{-2}>0$:
\begin{itemize}
\item If $\frac{d z_t}{d x}>0$, then the induced effects through $z_t$ satisfy
\[
\left.\frac{d s_{\phi,t}}{d x}\right|_{z\text{-channel}}<0,
\qquad
\left.\frac{d s_{L,t}}{d x}\right|_{z\text{-channel}}<0,
\qquad
\left.\frac{d s_{\pi,t}}{d x}\right|_{z\text{-channel}}>0.
\]
\item If $\frac{d z_t}{d x}<0$, these inequalities reverse.
\end{itemize}

\subsection{Direct effects for \texorpdfstring{$x\in\{\mu,\nu\}$}{x in {mu,nu}}}

Holding $z_t$ fixed, the direct partial derivatives are
\begin{align}
\left.\frac{\partial s_{\phi,t}}{\partial \mu}\right|_{z_t}
&= \frac{\nu}{\mu^2} z_t^{-1} > 0,
&
\left.\frac{\partial s_{\phi,t}}{\partial \nu}\right|_{z_t}
&= -\frac{1}{\mu} z_t^{-1} < 0,
\\
\left.\frac{\partial s_{L,t}}{\partial \mu}\right|_{z_t}
&= -\frac{\nu}{\mu^2}
\Big[(1-\alpha)-z_t^{-1}\Big],
&
\left.\frac{\partial s_{L,t}}{\partial \nu}\right|_{z_t}
&= \frac{1}{\mu}
\Big[(1-\alpha)-z_t^{-1}\Big],
\\
\left.\frac{\partial s_{\pi,t}}{\partial \mu}\right|_{z_t}
&= \frac{\nu}{\mu^2}
\Big(1-z_t^{-1}\Big) > 0,
&
\left.\frac{\partial s_{\pi,t}}{\partial \nu}\right|_{z_t}
&= -\frac{1}{\mu}
\Big(1-z_t^{-1}\Big) < 0.
\end{align}
In particular, the sign of the direct effect on the labour share $s_{L,t}$ depends on whether $z_t^{-1}$ is larger or smaller than $(1-\alpha)$.
\end{document}